\documentclass[12pt]{article}

\newcommand{\blind}{0}

\usepackage{amsmath}
\usepackage{graphicx,psfrag,epsf}
\usepackage{enumerate}
\usepackage{url} 
\usepackage{amsfonts}
\usepackage{diagbox} 
\usepackage{array,multirow,graphicx} 
\usepackage{bbm} 
 
\usepackage[a4paper]{geometry}

\usepackage[utf8]{inputenc} 
\usepackage[T1]{fontenc}
\usepackage{natbib}

\usepackage[english]{babel}

\usepackage{tabularx}
\usepackage{rotating}
\usepackage{bm}

\usepackage{subcaption}

\newcommand{\reals}{\mathbb{R}} 
\newcommand{\matr}[1]{\mathbf{#1}} 
\newcommand{\norm}{\mathcal{N}}
\newcommand{\dd}{\mathrm{d}}

\begin{document}

\def\spacingset#1{\renewcommand{\baselinestretch}%
{#1}\small\normalsize} \spacingset{1}


\if0\blind
{
  \title{\Large Spatially Varying Anisotropy for Gaussian Random Fields in Three-Dimensional Space}
  
  \author{\normalsize Martin Outzen Berild\thanks{Corresponding author, martin.o.berild@ntnu.no} and Geir-Arne Fuglstad \\
    \small Department of Mathematical Sciences,\\ \small Norwegian University of Science and Technology, Norway}
    \date{}
  \maketitle

} \fi
\if1\blind
{
  \bigskip
  \bigskip
  \bigskip
  \begin{center}
    {\LARGE\bf Spatially Varying Anisotropy for GRFs in 3D}
\end{center}
  \medskip
} \fi


\begin{abstract}

Isotropic covariance structures can be unreasonable for phenomena in
three-dimensional spaces such as the ocean. 
In the ocean, the variability of
the response may vary with depth, and ocean currents may lead to spatially
varying anisotropy. 
We construct a class of non-stationary anisotropic Gaussian random fields (GRFs) in three dimensions through stochastic partial differential equations (SPDEs) where computations are done using Gaussian Markov random field approximations. 

The approach is proven in a simulation study where the amount of data required to estimate these models is explored.
Then, the method is applied to construct a GRF prior on an ocean mass outside Trondheim, Norway, based on simulations from the complex numerical ocean model SINMOD. 
This GRF prior is compared to a stationary anisotropic GRF using in-situ measurements collected with an autonomous underwater vehicle where our approach outperforms the stationary anisotropic GRF for real-time prediction  of unobserved locations.

\end{abstract}

\noindent%

\vfill

\newpage
\spacingset{1.5} 

\section{Introduction}
Gaussian random fields (GRFs) are a powerful tool for spatial and spatio-temporal geostatistical modeling \citep{diggle1998model, cressie2015statistics}. When the key goal is predictions at unobserved locations, i.e., kriging, isotropic covariance functions often perform well, and more flexible covariance structures should be used with care \citep{fuglstad_does_2015}.
However, the screening effect in kriging \citep{stein2002screening} is not relevant in
other settings where the primary goal is the estimated covariance structure. E.g., to describe internal variability in a climate model ensemble \citep{castruccio2019reproducing}, or
to produce a spatial prior based on numerical simulations that will later be
used to guide autonomous sampling \citep{fossum2021learning,foss2021stSPDE}. 
For the former,  \citet{fuglstad2020compression,hu2021neuralSPDE} demonstrated that flexible covariance structures can perform better than stationary covariance structures.

There are many approaches to constructing flexible covariance structures \citep{nonStat_handbook2010,salvana2021lagrangian,schmidt2011considering}.
Some early approaches are the deformation method \citep{sampson1992nonparametric} and kernel convolutions \citep{paciorek2006spatial}, but
they both involve the covariances between 
any pair of locations. This means standard implementations are infeasible for large datasets. 
There are many ways to overcome such computational issues in spatial statistics and some are applicable for flexible covariance structures \citep{heatonEtAl2019}. The
stochastic partial differential equation (SPDE) approach \citep{lindgren_explicit_2011} is interesting because it directly gives rise to computationally efficient models and easily extends to non-stationary covariance models.

However, increasing the degree of flexibility in the covariance structure requires increasing the number of parameters. The common isotropic Matérn covariance functions \citep{stein2012interpolation} are parametrized through 3 parameters: marginal variance, range, and smoothness. Flexible models can have 100s or more parameters \citep{fuglstad_does_2015}.
An appealing way to reduce dimensionality is
to describe the covariance structure through covariates \citep{schmidt2011considering,neto2014accounting,ingebrigtsen2014spatial,ingebrigtsen2015estimation, risser2015regression}. 

The aforementioned works are all considering flexible covariance structures in two-dimensional space, and while the methods can be extended to three-dimensional space, the literature is sparse. For example, the SPDE approach has been used for simple anisotropic covariance structures in the context of fMRI data from the brain \citep{siden2019spatial}, and more complex covariance structures in the context of astronomy \citep{lee2021disks}, though this was two-dimensional space and time treated as three-dimensional space. However, spatially varying anisotropy in the SPDE approach \citep{fuglstad_exploring_2014} has not been extended to three-dimensional space.

The aim of this paper is to develop a new method for spatially varying anisotropy in three-dimensional space through the SPDE approach. A key advantage is that the formulation as an SPDE guarantees a valid covariance structure, and the main challenge is how to describe and parametrize non-stationary covariance structures. \citet{fuglstad_exploring_2014} used one vector field to describe spatially varying anisotropy, but in three dimensions, two spatially varying orthogonal vector fields are necessary for full generality.

In a simulation study, we investigate how much data is necessary to recover parameters for three different model complexities: stationary isotropic, stationary anisotropic, and non-stationary anisotropic. We then estimate GRF priors to encode knowledge about the ocean from a numerical forecast generated by the numerical model SINMOD by SINTEF. A stationary GRF prior and a non-stationary GRF prior are updated based on in-situ measurements by an autonomous underwater vehicle (AUV), and we evaluate the predictive ability during a mission in Trondheimsfjorden, Norway, on May 27, 2021.  Improved predictions are key, for example, in autonomous sampling of the oceans \citep{fossum2019toward, fossum2021learning}, but current approaches in autonomous ocean sampling are limited to stationary GRFs.

In Section 2, we describe how to model anisotropy and non-stationarity in three dimensions using SPDEs. Then in Section 3, we describe how to perform inference for the new model in a computationally efficient way. In Section 4, we describe the simulation study and discuss the results, and continue with the application to sampling in the ocean in Section 5. We end with a discussion in Section 6. 

\section{Constructing SPDEs with spatially varying anisotropy}
\label{sec:model}


\subsection{Existing models}
The Matérn covariance function on $\reals^3$ is given by
\begin{equation}
    r(\bm{s}_1,\bm{s}_2) = \frac{\sigma^2}{2^{\nu - 1}\Gamma(\nu)}(\kappa||\bm{s}_1-\bm{s}_2||)^\nu K_\nu (\kappa||\bm{s}_1-\bm{s}_2||),\quad \bm{s}_1, \bm{s}_2\in \reals^3,
    \label{eq:isoMatern}
\end{equation}
where $||\cdot||$ is the Euclidean distance in $\reals^3$,  $\sigma>0$ is the marginal standard deviation, $K_\nu$ is the modified Bessel function of the second kind and order $\nu>0$, and $\kappa>0$ is an inverse spatial scale parameter.  
As discussed in \citet{lindgren_explicit_2011},
GRFs with this covariance function is the stationary solutions of the SPDE
\begin{equation}
    (\kappa^2 - \nabla\cdot\nabla)^{\alpha/2}(\tau u(\bm{s})) = \mathcal{W}(\bm{s}), \hspace{25pt} \bm{s} \in \reals^3,
    \label{eq:isoSPDE}
\end{equation}
where $\alpha = \nu + 3/2$, $\tau = \sqrt{8\pi\kappa}/\sigma$, $\nabla\cdot\nabla$ is the Laplacian, and $\mathcal{W}$ is a standard Gaussian white noise process. 

 \citet{lindgren_explicit_2011} proposed to introduce non-stationarity by allowing $\kappa$ and $\tau$ to vary in space \citep{ingebrigtsen2014spatial,ingebrigtsen2015estimation} or by deformations of space \citep{hildeman2021deformed}. \citet{fuglstad_exploring_2014, fuglstad_does_2015} consider a version of the SPDE, where the Laplacian is replaced by an anisotropic Laplacian where the direction and degree of anisotropy vary spatially. This was further extended to spherical geometry in \citet{fuglstad2020compression,hu2021neuralSPDE}. However, all of these works were in two-dimensional base spaces, and only simpler models have been applied for three-dimensional base spaces \citep{siden2019spatial}.

The key idea in \citet{fuglstad_exploring_2014} was to replace $\nabla\cdot\nabla$ by $\nabla\cdot\mathbf{H}(\boldsymbol{s})\nabla$, where $\mathbf{H}(\boldsymbol{s})$ is everywhere a symmetric positive definite $2\times 2$ matrix that controls the strength and direction of anisotropy. The matrix-valued function was specified as $\mathbf{H}(\boldsymbol{s}) = \gamma(\boldsymbol{s})\mathbf{I}_2+\boldsymbol{v}(\boldsymbol{s})\boldsymbol{v}(\boldsymbol{s})^\mathrm{T}$, $\boldsymbol{s}\in\reals^2$, where $\gamma(\cdot)$ is a positive function and $\boldsymbol{v}(\cdot)$ is a vector field. This allows $\gamma(\cdot)$ to control the baseline strength of dependence in all directions, and $\boldsymbol{v}(\cdot)$ to control the strength and direction of additional spatial dependence. However, the same parametrization in $\reals^3$ is not sufficiently general to control anisotropy fully.

\subsection{Stationary anisotropy in $\reals^3$}
\label{subsec:spde}

We follow the idea in \citet{fuglstad_exploring_2014} for $\reals^2$, and change the SPDE in Equation \eqref{eq:isoSPDE} to
\begin{equation}
    (\kappa^2 - \nabla \cdot \mathbf{H}
    \nabla)u(\bm{s}) = \mathcal{W}(\bm{s}), \enspace \bm{s} \in \reals^3,
    \label{eq:aniSPDE}
\end{equation}
where $\nabla\cdot\mathbf{H}\nabla$ is an anisotropic Laplacian and the symmetric positive definite $3\times 3$ matrix $\mathbf{H}$ controls the anisotropy. The parameter $\tau$ has been dropped since $\kappa$ and $\mathbf{H}$ together control both marginal variance and correlation.

\begin{figure}[!ht]
    \centering
    \includegraphics[width=0.7\textwidth]{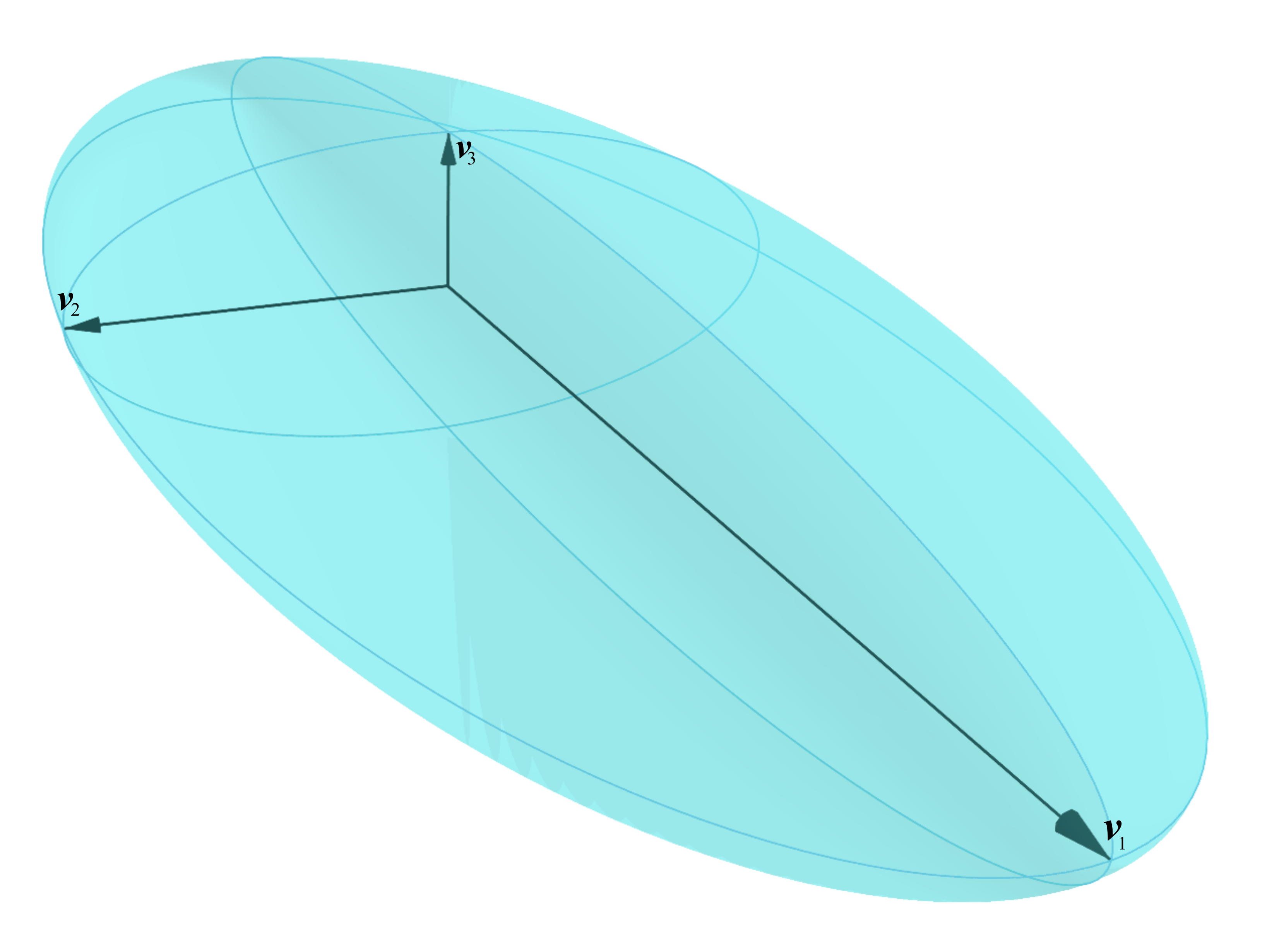}
    \caption{Iso-correlation surface at the $\sim$0.37 level of Equation~\eqref{eq:aniMatern2}, where $\bm{v}_1$, $\bm{v}_2$, and $\bm{v}_3$ are the eigenvectors of $\mathbf{H}$ with
    lengths $\sqrt{\lambda_1}/\kappa$, $\sqrt{\lambda_2}/\kappa$ and $\sqrt{\lambda_3}/\kappa$.
    \label{fig:corrH}}
\end{figure}

As shown in Appendix~A.1,
the resulting marginal variance is
\begin{equation}
    \sigma^2_m = \frac{1}{8\pi\kappa\sqrt{\det(\mathbf{H})}}
    \label{eq:margVar}
\end{equation}
and the covariance function is explicitly known as
\begin{equation}
    r(\bm{s}_1,\bm{s}_2) = \frac{1}{8\pi\kappa\sqrt{\det(\mathbf{H})}}\exp\left(-\kappa||\mathbf{H}^{-1/2}(\bm{s}_1-\bm{s}_2)||)\right)
    \label{eq:aniMatern2}
\end{equation}
for $\bm{s}_1, \bm{s}_2\in\reals^3$. The latter is derived in Appendix~A.2. This corresponds to geometric anisotropy in the Matérn covariance function with smoothness $\nu = 1/2$. To understand the behavior of the covariance function, it is useful to think about $\mathbf{H}$ in terms of its eigenvalue decomposition. Let
$\tilde{\bm{v}}_1$, $\tilde{\bm{v}}_2$, and $\tilde{\bm{v}_3}$ be orthonormal eigenvectors corresponding to eigenvalues $\lambda_1$, $\lambda_2$ and $\lambda_3$, respectively. Then 
Figure~\ref{fig:corrH} shows an example of the  0.37 level iso-correlation surface that will arise from the covariance function in Equation \eqref{eq:aniMatern2}.
The semi-axes of the ellipsoid in the figure are $\bm{v}_1 = (\sqrt{\lambda_1}/\kappa)\tilde{\bm{v}}_1$, $\bm{v}_2 = (\sqrt{\lambda_2}/\kappa)\tilde{\bm{v}}_2$, and $\bm{v}_3 = (\sqrt{\lambda_3}/\kappa)\tilde{\bm{v}}_3$, which by evaluating the covariance function with either of these semi-axes will yield the relationship and the iso-correlation level $r(\bm{v})/\sigma_m^2 = e^{-1} \approx 0.37$.

We generalize the parametrization described in Section \ref{subsec:spde} and $\mathbf{H}$ is decomposed as
\begin{equation}
    \mathbf{H} = \gamma \mathbf{I}_3 + \bm{v}\bm{v}^\mathrm{T} + \bm{\omega}\bm{\omega}^\mathrm{T}.
    \label{eq:parH}
\end{equation}
where $\bm{v} = (v_x,v_y,v_z)^\mathrm{T}\in\reals^3$ and $\bm{w} = (\omega_x,\omega_y,\omega_z)^\mathrm{T}\in\reals^3$, $\bm{v} \perp \bm{\omega}$, and $\gamma>0$. The eigenvalue decomposition of 
$\mathbf{H}$ has eigenvalues $\lambda_1 = \gamma$, $\lambda_2 = \gamma + ||\bm{v}||^2$ and $\lambda_3 = \gamma + ||\bm{w}||^2$ with the corresponding eigenvectors $\bm{v}_1=\bm{v} \times \bm{\omega}$, 
$\bm{v}_2=\bm{v}$ and $\bm{v}_3=\bm{\omega}$, respectively.
We construct $\bm{\omega}$ by a linear combination of two orthogonal vectors in the plane with $\bm{v}$ as normal vector. First, let $\bm{\omega}_1 = (-v_y, v_x, 0)^\mathrm{T}$, which satisfies
$\bm{v} \perp \bm{\omega}_1$.
Second, let 
$\bm{\omega}_2 = \bm{v}\times\bm{\omega}_1 = (-v_z v_x, -v_z v_y, v_x^2 + v_y^2)^\mathrm{T}$,
which also satisfies $\bm{v} \perp \bm{\omega}_2$. We parametrize $\bm{\omega}$ through
\begin{equation}
 \bm{\omega} = \rho_1 \frac{\bm{\omega}_1}{||\bm{\omega}_1||} + \rho_2 \frac{\bm{\omega}_2}{||\bm{\omega}_2||},
\end{equation}
where $\rho_1, \rho_2 \in\reals$ which works whenever $v_x = v_y \neq 0$. An alternative solution is to use Euler-Rodrigues parametrization \citep{euler_problema_1771,Rodrigues1840} to obtain both $\bm{v}$ and $\bm{\omega}$; however, in this case, the parameters are less interpretable and the issue is simply nullified by numerical optimization with appropriate initial parameter values.

The above parametrization  for $\mathbf{H}$ uses six parameters, $\gamma$, $v_x$, $v_y$, $v_z$, $\rho_1$, and $\rho_2$, to describe all forms of geometric anisotropy. The parameterization is interpretable: 1) $\gamma$ controls the isotropic effect, 2) $v_x$, $v_y$, and $v_z$ controls one anisotropy in one direction, and 3) $\rho_1$ and $\rho_2$ controls anisotropy in a second direction orthogonal to the first. Lastly, $\kappa$ simultaneously controls scaling of spatial dependence equally in all directions, and the variance of the GRF together with the six other parameters as seen in Equation \eqref{eq:margVar}.

\subsection{Spatially varying anisotropy on bounded domain $\mathcal{D}\subset\reals^3$}
Non-stationarity and spatially varying anisotropy is achieved by making the coefficients in Equation \eqref{eq:aniSPDE} spatially varying,
\begin{equation}
    (\kappa(\bm{s})^2 - \nabla \cdot \mathbf{H}(\bm{s})
    \nabla)u(\bm{s}) = \mathcal{W}(\bm{s}), \quad \bm{s} \in \reals^3,
    \label{eq:nonStatSPDEfull}
\end{equation}
where $\kappa(\cdot)$ is a positive function, and 
$\mathbf{H}$ is a spatially varying symmetric positive definite $3 \times 3$ matrix.
Heuristically, one can imagine that the SPDE is gluing together different local behavior described by ellipsoids, as discussed in Section \ref{subsec:spde}, to a valid non-stationary covariance structure.

In practice, we need to limit Equation \eqref{eq:nonStatSPDEfull} to a bounded domain to parametrize the non-stationarity. The SPDE we propose is 
\begin{equation}
    (\kappa(\bm{s})^2 - \nabla \cdot \mathbf{H}(\bm{s})
    \nabla)u(\bm{s}) = \mathcal{W}(\bm{s}), \quad \bm{s} \in \mathcal{D}\subset\reals^3,
    \label{eq:nonStatSPDE}
\end{equation}
where $\mathcal{D}$ is bounded, and we enforce the boundary condition 
$$(\mathbf{H}(\bm{s})\nabla u(\bm{s}))^\mathrm{T}\bm{n}(\bm{s}), \quad \bm{s}\in\partial \mathcal{D},$$
where $\bm{n}(\bm{s})$ is the outward normal vector of $\mathcal{D}$. This corresponds to no flux through the boundary. The effect of the boundary conditions is increased marginal variance on the boundary and increased spatial dependency due to the ``reflective'' boundary condition. As discussed in \citet{lindgren_explicit_2011,fuglstad_does_2015}, one can extend the domain $\mathcal{D}$ outside the area with observations to reduce boundary effects, or one can consider the boundary effects a feature that the non-stationary model can adjust for if necessary.

\section{Estimating SPDEs with spatially varying anisotropy}
\label{sec:estimation}

\subsection{Parameterizing the non-stationarity}
\label{subsec:nonstat}

Before using the SPDE in Equation \eqref{eq:nonStatSPDEfull} in inference, we parametrize the non-stationarity through a finite number of parameters. This involves expanding $\log(\kappa(\cdot))$, $\log(\gamma(\cdot))$, $v_x(\cdot)$, $v_y(\cdot)$, $v_z(\cdot)$, $\rho_1(\cdot)$, and $\rho_2(\cdot)$ in basis functions. The log-transform is used for $\kappa(\cdot)$ and $\gamma(\cdot)$ since they must be positive functions.

Let $g:\reals^3\rightarrow\reals$ denote a generic function that we want to expand in a basis, and let $p > 0$ the number of basis functions.
We use basis splines similar to \citet{fuglstad_does_2015}, and set
\begin{equation}
    g(\bm{s})  = \bm{f}(\bm{s})^\mathrm{T}\bm{\alpha}_{g},
    \label{eq:lincombBS}
\end{equation}
where $\bm{\alpha}_{g}\in\reals^p$, and $\bm{f}(\bm{s}) = (f_1(\bm{s}), \ldots, f_p(\bm{s}))^\mathrm{T}$ is a $p$-dimensional vector with the basis functions evaluated at location $\bm{s}$.

In this paper, we will use rectangular domains $\mathcal{D} = [A_1, B_1]\times[A_2, B_2]\times[A_3, B_3]$, and a basis constructed as 
a tensor product of three one-dimensional B-splines. This means that $p = m^3$, where $m > 0$ is the number of basis functions used in each dimension. We use clamped splines where the derivative is 0 at each boundary, and the construction of the clamped one-dimensional B-splines is discussed in Appendix~A.3. Figure~\ref{fig:bspline1d} shows an example of the resulting basis functions in 1-dimension. 

\begin{figure}[!htb]
    \centering
    \includegraphics[width=0.8\textwidth]{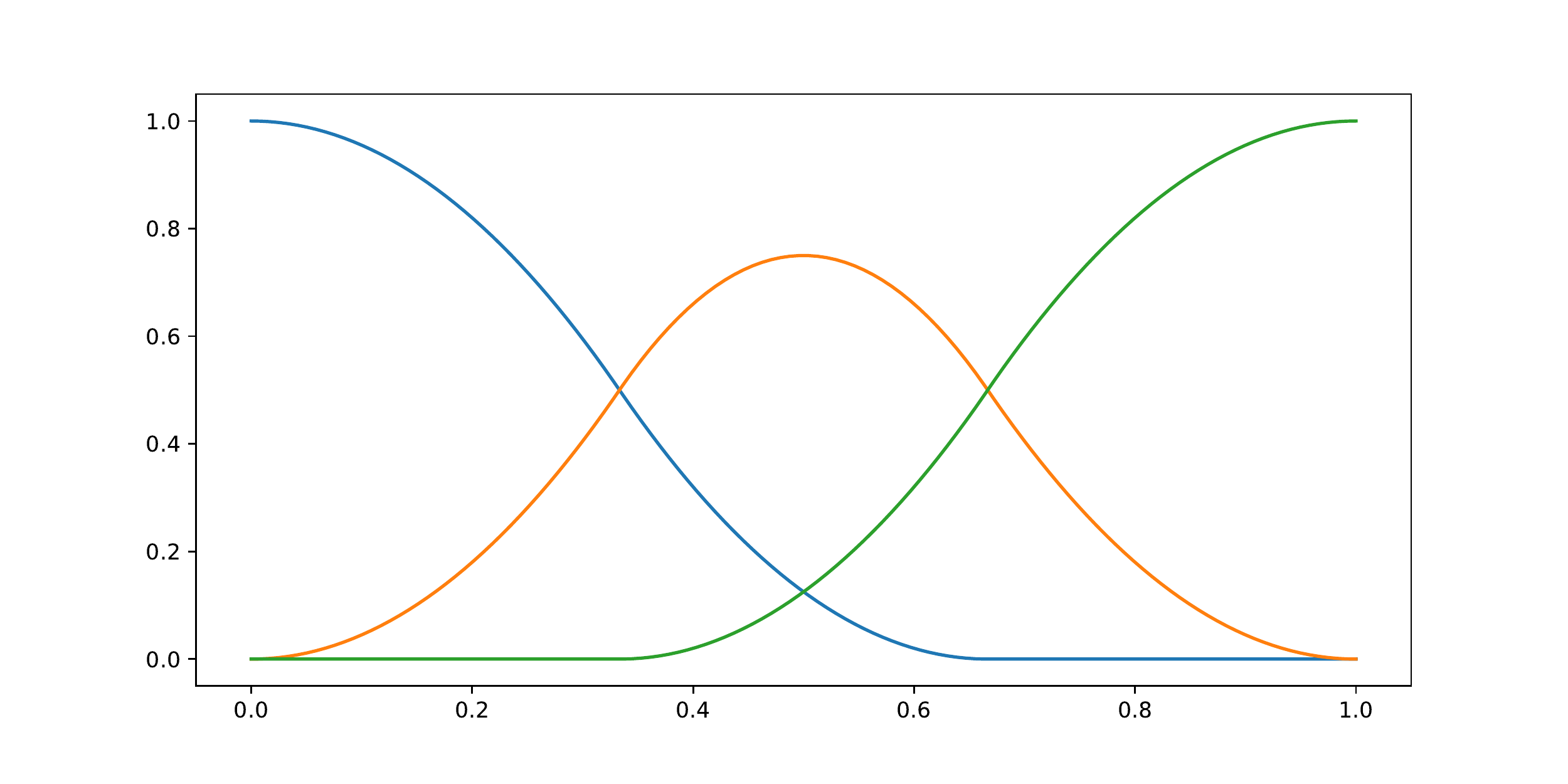}
    \caption{Clamped B-spline basis with three basis functions in 1D.}
    \label{fig:bspline1d}
\end{figure}

Let $B_{x,i}$ denote the $i$-th basis function of the second-order basis in the $x$-dimension, and similarly $B_{y,j}$ and $B_{z,k}$ for the $y$- and $z$-dimension. The resulting tree-dimensional basis is then
\begin{equation}
    f_{ijk}\left(\bm{s}\right)  = B_{x,i}(s_1)\cdot B_{y,j}(s_2) \cdot B_{z,k}(s_3), \quad \bm{s} = (s_1, s_2, s_3)^\mathrm{T}\in\mathcal{D},
    \label{eq:tensorBS}
\end{equation}
for all combinations $i,j,k \in \{1,\ldots, m\}$. This means that $\bm{\alpha}_g\in\reals^{m^3}$, and $m^3$ parameters must be estimated for each of the seven functions described at the start of the section.

\begin{figure}[!htb]
    \centering
    \includegraphics[width=0.5\textwidth]{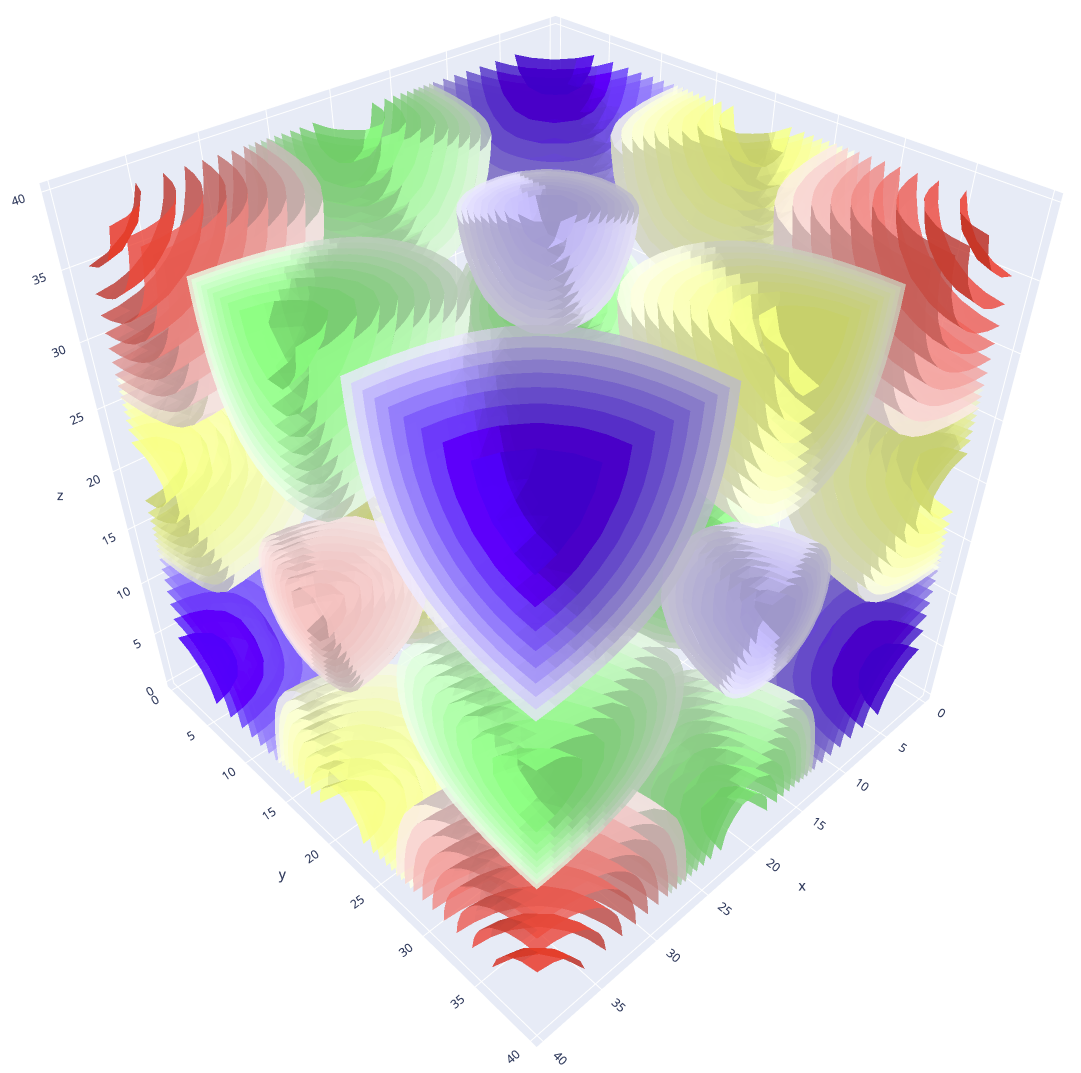}
    \caption{Parameterized function representation with B-splines in 3D.}
    \label{fig:bspline3d}
\end{figure}

In Sections \ref{sec:simstud} and \ref{sec:application}, we use $p = m^3 = 3^3 = 27$. For a total of 189 parameters in the seven functions. When data is sparse, such a model can easily result in overfitting \citep{fuglstad_does_2015}, and it is necessary to introduce penalties on the seven functions. In \citet{fuglstad_does_2015}, this was achieved by a hierarchical model where
\[
    \tau_g \Delta g(\boldsymbol{s}) = \mathcal{W}_g(\boldsymbol{s}), \quad \bm{s}\in\mathcal{D},
\]
together with Neumann boundary conditions of zero derivatives on the boundary of the domain. However, this requires selecting a reasonable value for $\tau_g > 0$ for each of the seven functions and is computationally expensive if it is done using cross-validation. However, in the context of this paper, we are constructing a stochastic model that mimics the behavior of a densely ``observed'' numerical simulation model and does not include penalties beyond the restriction of using 27 basis functions. We demonstrate the ability of this model to be estimated in our context in the simulation study in Section \ref{sec:simstud}, and also investigate the amount of data needed to estimate the model.

\subsection{Hierarchical model and discretization}
\label{subsec:hierarch}
Consider a bounded domain $\mathcal{D}\subset\reals^3$, and 
observations $\bm{y} = (y_1,y_2,\dots,y_n)$ made at locations $\bm{s}_1,\bm{s}_2,\dots,\bm{s}_n \in \mathcal{D}$. We assume a Gaussian observation model 
\[
    y_i|\eta(\bm{s}_i), \sigma_\mathrm{N}^2 \sim \mathcal{N}(\eta(\bm{s}_i), \sigma_\mathrm{N}^2), \quad i = 1, \ldots, n,
\]
where $\sigma_\mathrm{N}^2> 0$ is the nugget variance and 
\[
    \eta(\bm{s}) = \boldsymbol{x}(\bm{s})^\mathrm{T}\bm{\beta} + u(\bm{s}), \quad \bm{s}\in\mathcal{D},
\]
describes true spatial variation as a combination of covariates and a GRF. Here $\boldsymbol{x}(\cdot)$ is a spatially varying vector of $k$ covariates, $\bm{\beta}\in\reals^k$ are the coefficients of the covariates, and $u(\cdot)$ is a GRF with spatially varying anisotropy as presented in Section \ref{sec:model}.

As described in Appendix~B, the GRF $u(\cdot)$ is discretized using a regular grid with $l$ cells, and we get a Gaussian Markov random field $\bm{w} = (w_1, \ldots, w_l)^\mathrm{T}$. Let $\bm{\theta}$ be the vector of all parameters controlling $u(\cdot)$, then
\[
    \bm{w}|\boldsymbol{\theta} \sim \mathcal{N}_l(\bm{0}, \mathbf{Q}^{-1}),
\]
where dependence on $\bm{\theta}$ is suppressed for $\mathbf{Q}$, and $\mathbf{Q}$ is a $l\times l$ precision matrix with a three-dimensional spatial sparsity structure. The vector $\bm{w}$ is linked to $u(\cdot)$ through a linear transformation $u(\bm{s}) = \bm{a}(\bm{s})^\mathrm{T}\bm{w}$, where $\bm{a}$ has only one non-zero entry corresponding to which grid cell location $\bm{s}$ belongs. This gives $\bm{u} = (u(\bm{s}_1), \ldots, u(\bm{s}_n))^\mathrm{T} = \mathbf{A}\bm{w}$, where the $n\times l$ matrix $\mathbf{A}$ only has one non-zero entry on each row.
 
The coefficients of the fixed effect, $\bm{\beta}$, is assigned the weak penalty $\bm{\beta} \sim \mathcal{N}_K(\bm{0},V\mathbf{I}_K)$ for a fixed $V>0$. Thus we can write $\bm{y}$ as
\begin{equation}
    \bm{y} =  \mathbf{X}\bm{\beta} + \mathbf{A}\bm{w} + \bm{\epsilon},
    \label{eq:modely}
\end{equation}
where $\mathbf{X}$ is the design matrix of covariates, and $\bm{\epsilon} \sim \mathcal{N}_n(\bm{0},\mathbf{I}_n \sigma_\mathrm{N}^2)$ is an $n$-dimensional vector of random noise. This gives rise to the hierarchical formulation
\begin{align*}
    \bm{y}|\bm{\beta}, \bm{w}, \sigma_\mathrm{N}^2 \sim \mathcal{N}_n(\mathbf{X}\bm{\beta}+\mathbf{A}\bm{w}, \sigma_\mathrm{N}^2\mathbf{I}_n), \\
    \bm{\beta}\sim\mathcal{N}_k(\bm{0}, V\mathbf{I}_k), \quad \bm{w}|\bm{\theta}\sim\mathcal{N}_l(\bm{0}, \mathbf{Q}^{-1}).
\end{align*}

Let $\bm{s}^*\in\mathcal{D}$ be an unobserved location.
After parameters $\hat{\bm{\theta}}$ and $\hat{\sigma_\mathrm{N}^2}$ are estimated, one can predict the underlying value $\eta(\bm{s}^*) = \boldsymbol{x}(\bm{s}^*)^\mathrm{T}\bm{\beta} + \bm{a}(\bm{s}^*)^\mathrm{T}\bm{w}$ or a new observation $y^* = \boldsymbol{x}(\bm{s}^*)^\mathrm{T}\bm{\beta} + \bm{a}(\bm{s}^*)^\mathrm{T}\bm{w}+ \epsilon^*$, where $\epsilon^*\sim \mathcal{N}(0, \hat{\sigma_\mathrm{N}^2})$ is a new nugget. The predictions are made using the conditional distributions $\eta(\bm{s}^*)|\bm{y}, \bm{\theta} = \hat{\bm{\theta}}, \sigma_\mathrm{N}^2 = \hat{\sigma_\mathrm{N}^2}$ and $y^*|\bm{y}, \bm{\theta} = \hat{\bm{\theta}}, \sigma_\mathrm{N}^2 = \hat{\sigma_\mathrm{N}^2}$. The estimation of parameters is detailed in the next section.

\subsection{Parameter inference}
\label{subsec:paraminf}
Simplify notation by letting $\bm{z} = (\bm{u}^\mathrm{T},\bm{\beta}^\mathrm{T})^\mathrm{T}$. Then
\begin{equation*}
    \bm{z} | \bm{\theta}  \sim \mathcal{N}(\bm{0}, \mathbf{Q}_z^{-1}), \enspace \textrm{where } \mathbf{Q}_z = \begin{bmatrix}
    \mathbf{Q} & \mathbf{0} \\
    \mathbf{0} & V\mathbf{I}_k
    \end{bmatrix}.
\end{equation*}
Let $\mathbf{S}= \begin{bmatrix} \mathbf{A} & \mathbf{X}\end{bmatrix}$, then the observation model can be
rewritten as
\begin{equation}
    \bm{y} | \bm{z}, \sigma_\mathrm{N}^2 \sim \mathcal{N}_n(\mathbf{S}\bm{z},\mathbf{I}_n \sigma_\mathrm{N}^2).
    \label{eq:obsmod}
\end{equation}
Using this notation the log-likelihood can be expressed as
\begin{equation}
    \begin{aligned}
    \log \pi(\bm{\theta}, \sigma_\mathrm{N}^2|\bm{y}) = & \textrm{  Const}  + \log \pi(\bm{\theta}, \sigma_\mathrm{N}^2) + \frac{1}{2}\log\det\left(\mathbf{Q}_z\right) - \frac{n}{2} \log(\sigma_\mathrm{N}^2) \\
                                   & - \frac{1}{2}\log\det\left(\mathbf{Q}_\mathrm{C}\right)- \frac{1}{2}\bm{\mu}_\mathrm{C}^\mathrm{T}\mathbf{Q}_\mathrm{C}\bm{\mu}_\mathrm{C}
                                   - \frac{1}{2\sigma_\mathrm{N}^2}(\bm{y}-\mathbf{S}\bm{\mu}_\mathrm{C})^\mathrm{T}(\bm{y}-\mathbf{S}\bm{\mu}_\mathrm{C}).
    \end{aligned}
    \label{eq:posterior}
\end{equation}
Here dependence on $\bm{\theta}$ is suppressed for $\bm{\mu}_\mathrm{C}$, $\mathbf{Q}_z$ and $\mathbf{Q}_\mathrm{C}$, and $\pi(\bm{\theta}, \sigma_\mathrm{N}^2)$ can be used to assign a penalty on $\bm{\theta}$, e.g., like the random-walk penalty used in \citet{fuglstad_does_2015}. The conditional precision matrix $\mathbf{Q}_\mathrm{C}$ is 
\begin{equation}
\mathbf{Q}_\mathrm{C} = \mathbf{Q}_z + \mathbf{S}^\mathrm{T}\mathbf{S}/\sigma_\mathrm{N}^2
\label{eq:condcov}
\end{equation}
and $\bm{\mu}_\mathrm{C}$ is the conditional mean,
\begin{equation}
    \bm{\mu}_\mathrm{C}=\mathbf{Q}_\mathrm{C}^{-1}\mathbf{S}^\mathrm{T}\bm{y}/\sigma_\mathrm{N}^2.
    \label{eq:condmean}
\end{equation}

Parameter inference is done by maximizing Equation \eqref{eq:posterior} with respect to $\bm{\theta}$ and $\sigma_\mathrm{N}^2$.
The parameter vector $\bm{\theta}$ includes all coefficients for the basis functions, and when using 27 basis functions for each function,
\begin{equation*}
    \bm{\theta} = \left(\bm{\alpha}_{\log (\kappa^2)},\bm{\alpha}_{\log \gamma},\bm{\alpha}_{v_x},\bm{\alpha}_{v_y},\bm{\alpha}_{v_z},\bm{\alpha}_{\rho_1}, \bm{\alpha}_{\rho_2}\right),
\end{equation*}
has 189 parameters. 
The parameter space is challenging to search
and we use an analytical expression for the gradient in the optimization algorithm.  
The derivation of the analytical gradient involves many nested chain rules and a technique to calculate a partial inverse of sparse matrices \citep{rue_markov_2010}, see Appendix~A.5 for a complete description.

\section{Simulation study}
\label{sec:simstud}
In this section, we perform a simulation study to investigate the amount of data required to acquire reasonable parameter estimates of models with varying complexity that are specified through the SPDE. A comparison of these estimates is made from simulated data generated from three different parametrizations of the covariance structures. 

The observation model for the different parametrizations is
\begin{equation}
    \bm{y}_{\mathrm{mod}} = \matr{A}\bm{w}_{\mathrm{mod}} + \bm{\epsilon},
    \label{eq:simstudmod}
\end{equation}
where $\bm{w}_{\mathrm{mod}}$ is the GMRF controlled by the parameters $\bm{\theta}_\mathrm{mod}$ in the respective models, and $\bm{\epsilon}$ is the independent noise term with mean zero and standard deviation $\bm{\sigma}_N = 0.1$ which is identical for all the parametrizations. 
Furthermore, the models are discretized on the same domain with a grid of size $(M, N, P) = ( 30, 30, 30)$ resulting in a total of 27000 grid nodes where the center of which is our spatial locations $\bm{s}\in\mathcal{D} = [A_1, B_1]\times[A_2, B_2]\times[A_3, B_3] = [0,40]\times[0,40]\times[0,40]$.

The first and simplest model is a Stationary Isotropic (SI) model which has a covariance structure controlled by the three parameters $\bm{\theta}_\mathrm{SI} = (\log \kappa^2, \log \gamma, \log \sigma_N^2)$, that is assigned to the values $\kappa^2 = 0.2$, $\gamma=2.5$ and $\sigma_N = 0.1$. The resulting spatial range is 10.59 with a marginal variance of 0.023.

The second is a Stationary Anisotropic (SA) model composed of the 8 parameters $\bm{\theta}_\mathrm{SA} = (\log \kappa^2, \log \gamma, v_x, v_y, v_z, \rho_1, \rho_2, \log \sigma_N^2)$ set to $\kappa^2= 0.35$, $\gamma = 0.5$, $v_x = 1.9$, $v_y=1.4$, $v_z = 0.4$, $\rho_1 = 1.4$, $\rho_2 = 0.6$ and $\sigma_N = 0.1$. This results in spatial ranges of 10.08 along the $x$-dimension, 6.75 along $y$, and 3.88 along $z$ with a marginal variance of 0.023. 

\begin{figure}[!ht]
    \centering
    \begin{subfigure}[b]{0.45\textwidth}
         \centering
         \includegraphics[width=\textwidth]{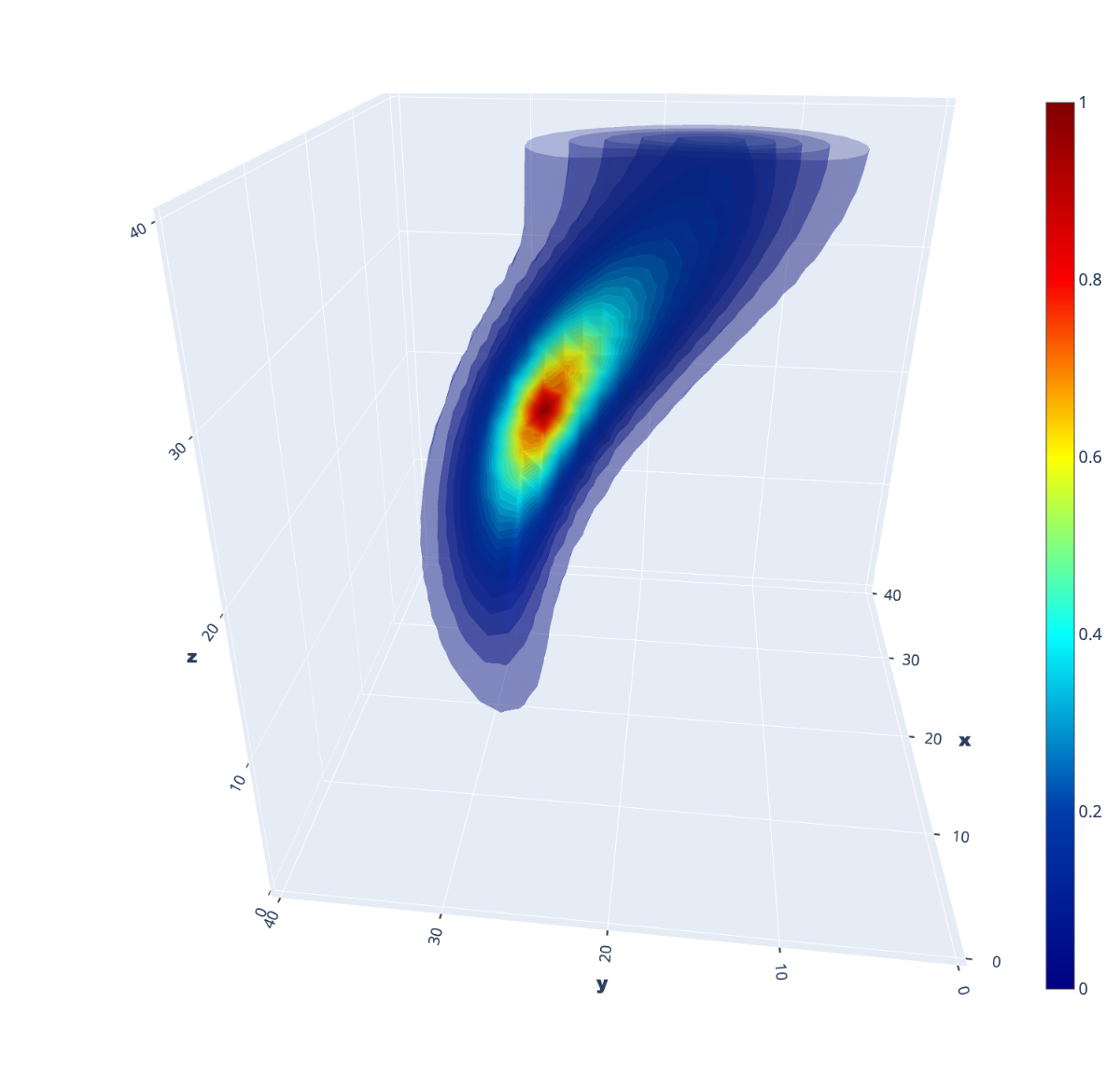}
         \caption{Correlation}
    \end{subfigure}
    \begin{subfigure}[b]{0.45\textwidth}
         \centering
         \includegraphics[width=\textwidth]{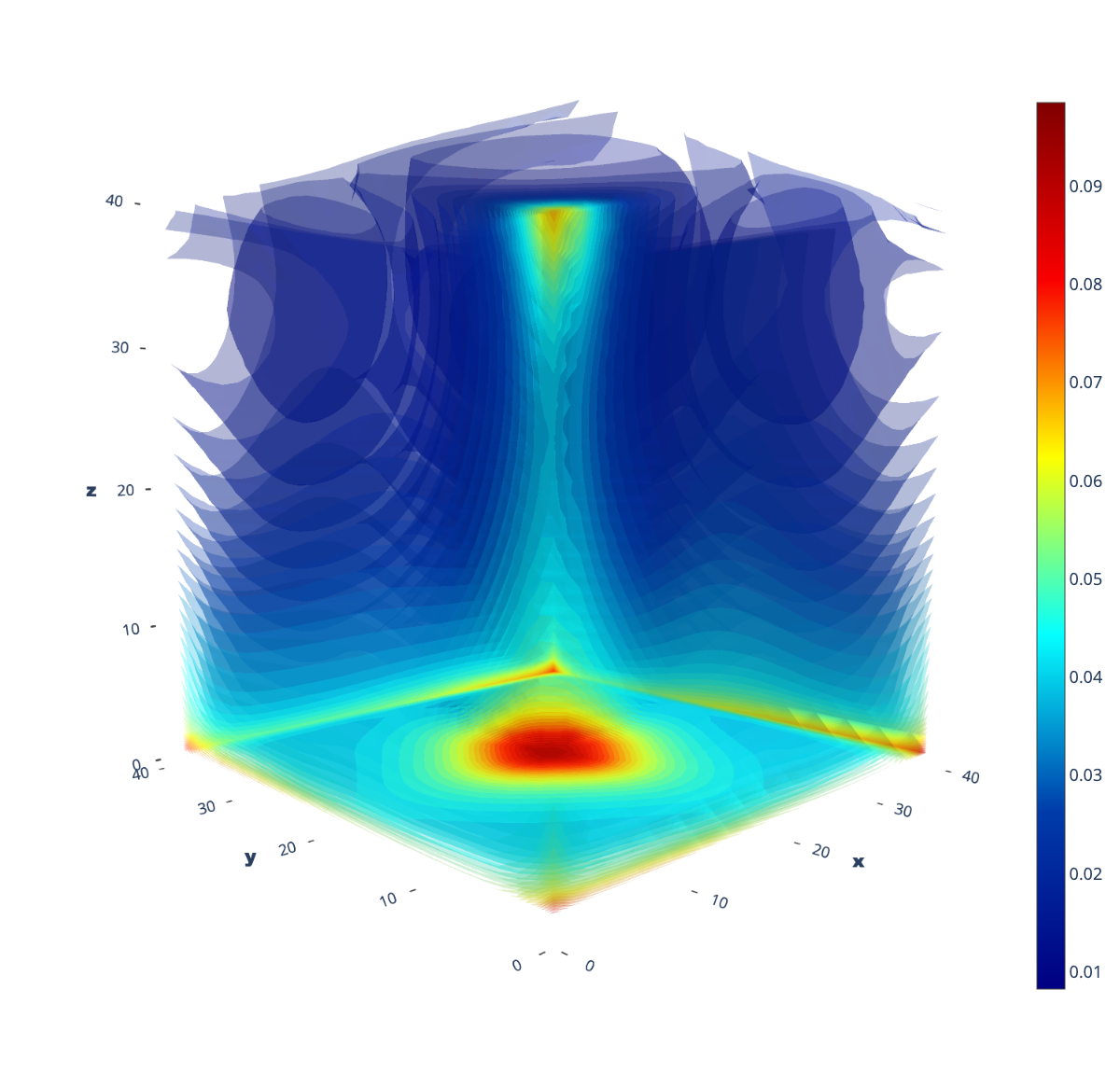}
         \caption{Marginal Variance}
    \end{subfigure}
    \caption{\label{fig:NA_simstud}Spatial correlation at location [26,26,20] (a) and variance of the spatial effect (b) in the non-stationary anisotropic model.}
\end{figure}

The parameters of these first two models are simply assigned some reasonable value; however, the third and most complex model with a non-stationary anisotropic covariance and a total of 190 parameters, they are much more troublesome to select. Therefore, functions are chosen to assign the parameter values in $\bm{\theta}_\mathrm{NA}$ throughout the domain $\mathcal{D}$ such that the dependency directions imitate a vortex. Using these functions and evaluating them at the spatial locations in the discretization the parameters of the B-splines, described in Section~\ref{subsec:nonstat}, are found by optimization. These aforementioned parameters are  $\bm{\theta}_\mathrm{NA} = \left(\bm{\alpha}_{\log (\kappa^2)},\bm{\alpha}_{\log \gamma},\bm{\alpha}_{v_x},\bm{\alpha}_{v_y},\bm{\alpha}_{v_z},\bm{\alpha}_{\rho_1}, \bm{\alpha}_{\rho_2},  \log \sigma_N \right)$ with $\sigma_N = 0.1$, and the resulting covariance structure can be viewed in Figure~\ref{fig:NA_simstud}.

We will now examine the extent of data required to fit back the parameters of the three models described above. 
First, we simulate multiple datasets from the observation model, Equation~\eqref{eq:simstudmod}, with a different number of observed spatial locations and realizations (replicated observations of these spatial locations).
The number of spatial locations varies between 100, 10000, and 27000 (all), and the number of realization range between 1, 10, and 100, so nine different combinations of dataset sizes. Furthermore, we want to perform 100 different trials for each of these combinations, and thereby have 900 total datasets per model. Also, note that the observed spatial locations are randomly chosen in each trial. From this, some statistics can be recovered about the model estimates that can give insight into the applicability of the different parameterizations. 

\setlength\tabcolsep{4.0pt} 
\begin{table}
\caption{The Root Mean Square Error (RMSE) of parameter estimates in the stationary isotropic, stationary anisotropic, and non-stationary anisotropic model from 100 independent trials for each combination of dataset sizes; the number of observed locations (No. loc.) and the number of replicated observations of these locations (No. real.).}
\small
\label{tab:simstud1}
\centering
\begin{tabular}{|c|c||c|c|c|c|c|c|c|c|c|}
\hline
 \multicolumn{2}{|c||}{No. loc.} & \multicolumn{3}{c|}{100}  & \multicolumn{3}{c|}{10000}& \multicolumn{3}{c|}{27000} \\
 \hline
\multicolumn{2}{|c||}{No. real.} & 1 & 10 & 100 & 1 & 10 & 100 & 1 & 10 & 100 \\
 \hline
 \cline{1-6}
 \parbox[t]{3mm}{\multirow{3}{*}{\rotatebox[origin=c]{90}{Stat. Iso.}}}
 &$\log \kappa$ & 0.763 & \textbf{0.168} & \textbf{0.047} & \textbf{0.123}  \\ 
 &$\log \gamma$ & 0.626 & \textbf{0.164} & \textbf{0.062} & \textbf{0.032} \\
 &  $\log \tau$ & 2.670 & \textbf{0.674} & \textbf{0.182} & \textbf{0.049} \\
 \cline{1-6}
 \cline{1-7}
 \parbox[t]{3mm}{\multirow{8}{*}{\rotatebox[origin=c]{90}{Stationary Anisotropic}}} 
 &$\log \kappa$ & 0.876 & 0.195 & \textbf{0.081} & \textbf{0.094} & \textbf{0.038} \\
 &$\log \gamma$ & 8.289 & 5.601 & \textbf{0.463} & \textbf{0.228} & \textbf{0.079} \\
 &      $|v_x|$ & 1.208 & 0.785 & \textbf{0.440} & \textbf{0.200} & \textbf{0.070} \\
 &      $|v_y|$ & 1.040 & 0.679 & \textbf{0.354} & \textbf{0.152} & \textbf{0.035} \\
 &      $|v_z|$ & 1.091 & 0.498 & \textbf{0.214} & \textbf{0.075} & \textbf{0.027} \\
 &   $|\rho_1|$ & 0.977 & 0.801 & \textbf{0.249} & \textbf{0.129} & \textbf{0.038} \\
 &   $|\rho_2|$ & 1.337 & 0.489 & \textbf{0.275} & \textbf{0.078} & \textbf{0.027} \\
 &  $\log \tau$ & 1.977 & 1.352 & \textbf{0.182} & \textbf{0.189} & \textbf{0.028} \\
 \cline{1-7}
 \cline{1-2} \cline{8-11}
 \parbox[t]{3mm}{\multirow{8}{*}{\rotatebox[origin=c]{90}{Non-Stationary Anisotropic}}}
 &$\log \kappa$ & \multicolumn{5}{c|}{} & 2.572 & 0.811 & \textbf{0.356} & \textbf{0.269}\\
 &$\log \gamma$ & \multicolumn{5}{c|}{} & 2.615 & 1.173 & \textbf{0.694} & \textbf{0.585}\\
 &      $|v_x|$ & \multicolumn{5}{c|}{} & 1.929 & 0.742 & \textbf{0.531} & \textbf{0.509}\\
 &      $|v_y|$ & \multicolumn{5}{c|}{} & 2.699 & 0.668 & \textbf{0.453} & \textbf{0.432}\\
 &      $|v_z|$ & \multicolumn{5}{c|}{} & 1.591 & 0.610 & \textbf{0.343} & \textbf{0.296}\\
 &   $|\rho_1|$ & \multicolumn{5}{c|}{} & 0.144 & 0.714 & \textbf{0.287} & \textbf{0.210}\\
 &   $|\rho_2|$ & \multicolumn{5}{c|}{} & 0.420 & 0.604 & \textbf{0.376} & \textbf{0.344}\\
 &  $\log \tau$ & \multicolumn{5}{c|}{} & 1.152 & 0.017 & \textbf{0.005} & \textbf{0.005}\\
 \cline{1-2} \cline{8-11}
\end{tabular}
\end{table}
\setlength\tabcolsep{6pt} 

Table~\ref{tab:simstud1} shows the root mean square error (RMSE) between the set parameter values in each model and their values inferred by the different datasets. This was obtained using the inference method described in Section~\ref{subsec:paraminf} with the observation model in Equation~\eqref{eq:simstudmod} for each respective parametrization and trial. The columns describe the different number of observation locations (No. loc.) and the number of realizations (No. real.), and the different blocks represent the different models. The columns highlighted in bold for each respective model are the ones we have deemed as reasonable parameter estimates. Also, note that some parts of the table are omitted to simplify the presentation of the results for the reader as the full table does not affect the conclusion of this study.
From Table~\ref{tab:simstud1} we observe that the (simple) stationary models, SI and SA, require very little data. In fact, observing under 1\% of the grid for 10 realizations or more is good enough for the SI and the SA only requires some more realizations to attain similar parameter accuracy. 

On the other hand, the most flexible parameterization, the NA model, requires much more data and only reaches reasonable parameter accuracy when the whole grid is observed with 10 or more realizations. Now there is a large discrepancy between 10000 observed points (~37\%) and 27000 (100\%), so it could be interesting to investigate where in this range reasonable estimates are obtained. However, we have not chosen to explore this here. We also want to note that these estimates will change with the complexity of the covariance structure and with the initial values in the optimization. 

\section{GRF prior for statistical sampling of the ocean}
\label{sec:application}

\subsection{Aim}
Forecasts produced by numerical ocean models describe realistic behavior for the ocean, but local behavior such as plumes created by freshwater discharge from a river into the ocean are hard to accurately forecast. However, we can construct a prior based on the numerical ocean model that informs prior beliefs about the ocean, which can aid AUVs to more effectively sample the ocean. In this paper, the goal is to determine the three-dimensional extent of a freshwater plume in the ocean, and we assume operation time is short enough to justify a purely spatial prior that does not assume dynamical changes in time. 

There are two steps in our approach. Step 1 is to estimate a stationary GRF prior and a non-stationary GRF prior based on a simulation from the numerical ocean model as described in Section \ref{subsec:numstat}. Step 2 is to combine each of the estimated priors with an observation model, and evaluate the predictive ability on in-situ observations from AUV as described in Section \ref{subsec:datacollect}. The GRFs that we estimate based on the numerical ocean model can be viewed as statistical emulators of the ocean. 

\subsection{The numerical ocean model and the GRF prior}
\label{subsec:numstat}

The model training data used in this application is from a forecast produced by the ocean model SINMOD. Data is provided by SINTEF Ocean which developed and ran the simulation. SINMOD is a three-dimensional numerical ocean model based on primitive equations that are solved using finite difference methods on a regular grid with horizontal cell sizes of 20km$\times$20km and is nested in several steps down to 32m $\times$ 32m. Moreover, it uses z* vertical layers which allow for varying grid resolutions depending on the depth and help capture the higher variability of the surface. SINMOD is driven by atmospheric forces, freshwater outflows, and tides, and it provides numerical simulations of multiple variables such as salinity, temperature, and currents. The reader is referred to \citet{slagstad_modeling_2005} for a more detailed description of the method.

\begin{figure}[!ht]
    \centering
    \includegraphics[width=0.8\textwidth]{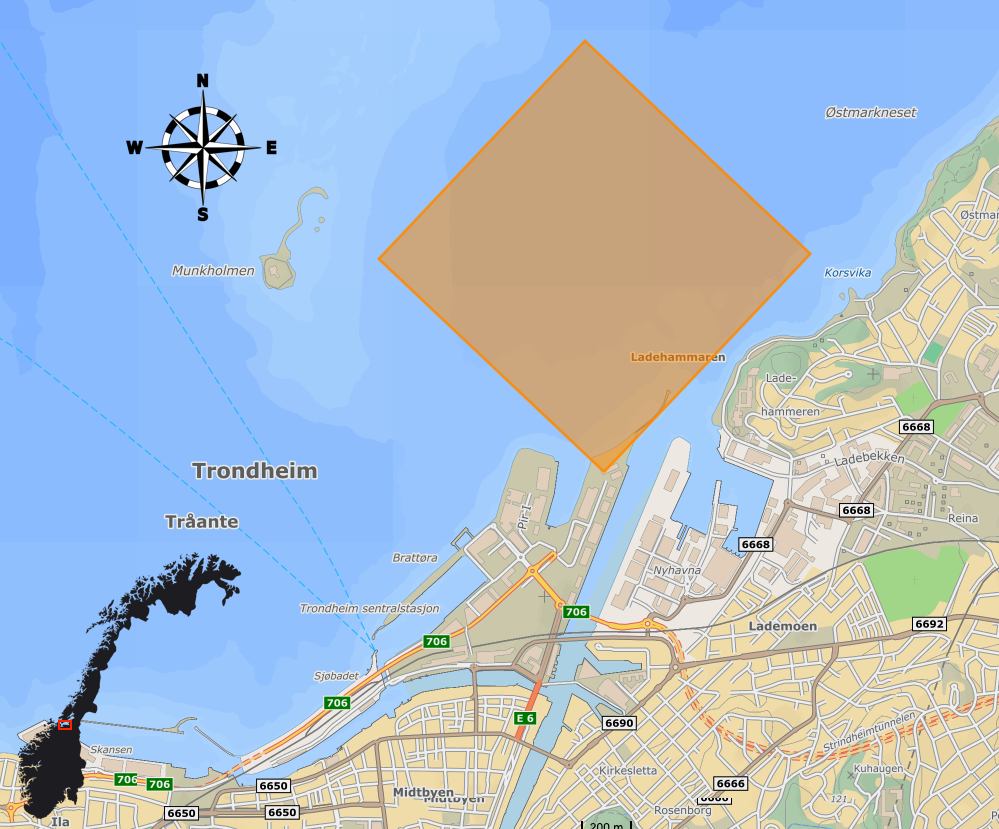}
    \caption{The area of operation in Trondheimsfjorden at Ladehammaren just outside of Trondheim, Norway. The compass shows the cardinal directions relative to the map.}
    \label{fig:AoO}
\end{figure}

The area of operation is located in Trondheimsfjorden at Ladehammaren just outside of Trondheim, Norway, and the operation date, the time measurements are collected with the AUV, is May 27, 2021, between 10:30 and 14:30. The outlined area in Figure~\ref{fig:AoO} indicates the operational area which covers 1408m $\times$ 1408m in the horizontal plane. 
At the southeast side of this field, the Nidelva river flows into the fjord. This causes a very dynamic salinity field that is unfeasible to describe with a stationary covariance model.
Therefore, we will use the numerical simulations from SINMOD to estimate a non-stationary GRF. As demonstrated in the simulation study, complex covariance structures can reliably be estimated based on such dense data.

\begin{figure}[!ht]
    \centering
    \includegraphics[width=0.99\textwidth]{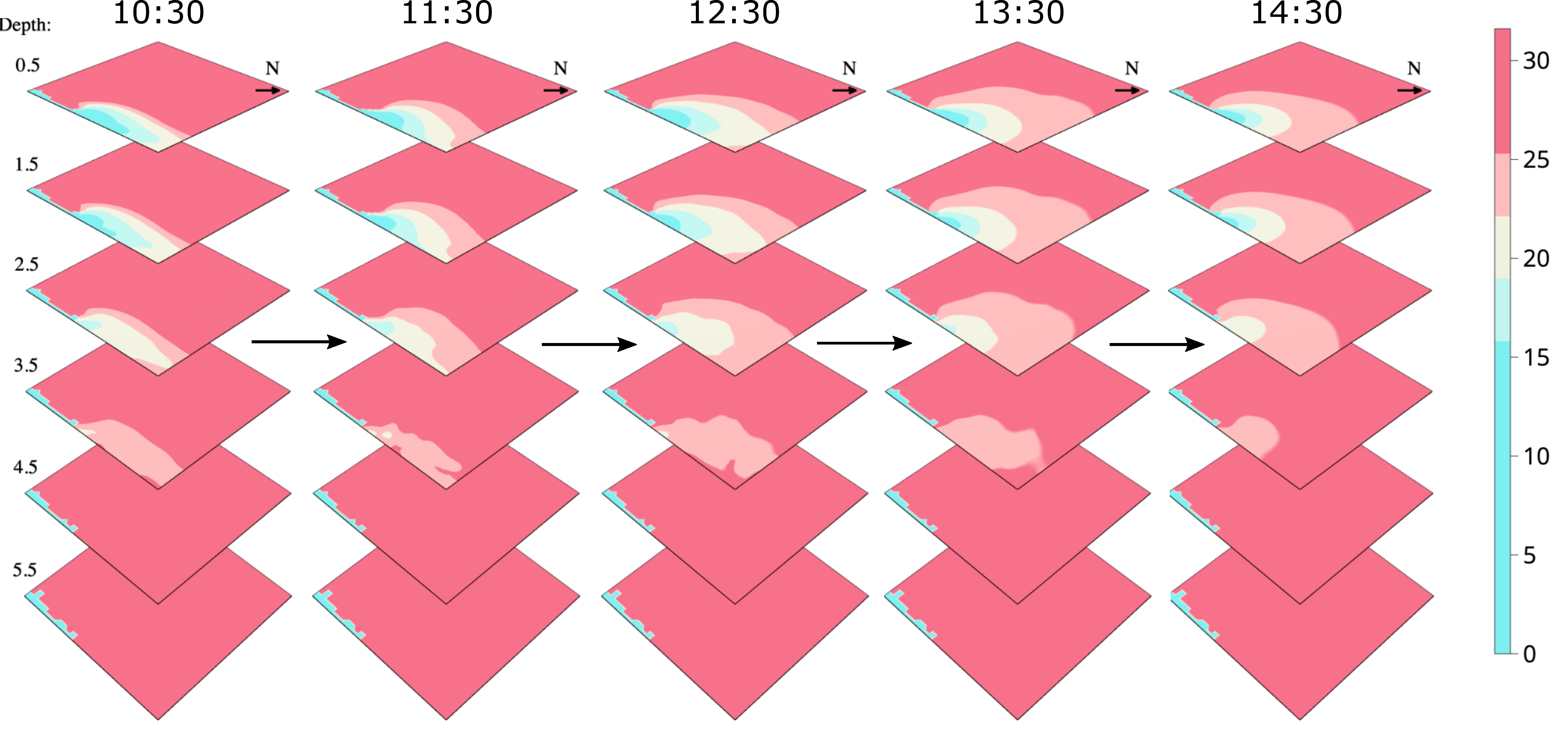}
    \caption{Five timesteps of the dataset simulated with the numerical ocean model SINMOD for May 27, 2021. The timestamps are displayed over their respective timesteps. The N-arrow is the cardinal north.}
    \label{fig:numsim}
\end{figure}

In this application, we will focus on univariate modeling of the salinity and we choose the fine-scale horizontal grid sizes $h_x = 32\,\mathrm{m}$  $h_y = 32\,\mathrm{m}$, which in total gives $N = 45$ and $M=45$ grid nodes for both the numerical and the statistical model. Moreover, in the vertical plane, we use 1-meter increments between the depth layers, i.e., $h_z = 1\,\mathrm{m}$. To avoid any major effects of the boundaries in this direction $P = 11$ depth layers are used resulting in a depth range of 0.5m to 10.5m. SINMOD outputs $\boldsymbol{z}_t$, $t = 0,1, 2, \ldots, 143$, which are vectors of salinity values in all cells in the three-dimensional grid at different time points throughout the whole May 27, 2021. The timesteps are 10 minutes, and Figure~\ref{fig:numsim} shows five timesteps from SINMOD for the top six depth layers during the operation.
Note that the varying vertical layers in the numerical model are either with 0.5m or 1m increments, so the SINMOD simulations don't require any additional modification to fit within our statistical model.

We first estimate the model 
\[
    \boldsymbol{z}_t = \Phi\boldsymbol{z}_{t-1} + \boldsymbol{\epsilon}_t, \quad t = 1, \ldots, 143,
\]
where $\Phi$ is a diagonal matrix of AR(1) coefficients. The diagonal entries of $\Phi$ are estimated with maximum likelihood separately for each spatial location such that
$\hat{\Phi}_{ii} = \sum_{t=1}^{143} z_{t,i}z_{t-1,i}/\sum_{t=1}^{143} z_{t-1,i}^2$ for $i = 1, \ldots, NMP$, where $z_{t,i}$ is the value in cell $i$ at time $t$.
We then compute empirical innovations $\hat{\boldsymbol{\epsilon}}_t = \boldsymbol{z}_t-\hat{\Phi}\boldsymbol{z}_{t-1}$, $t = 1, \ldots, 143$. These empirical innovations describe the spatial covariance structure for short-term changes in salinity.

\begin{figure}[!ht]
    \centering
    \begin{subfigure}[b]{0.3\textwidth}
         \centering
         \includegraphics[width=\textwidth]{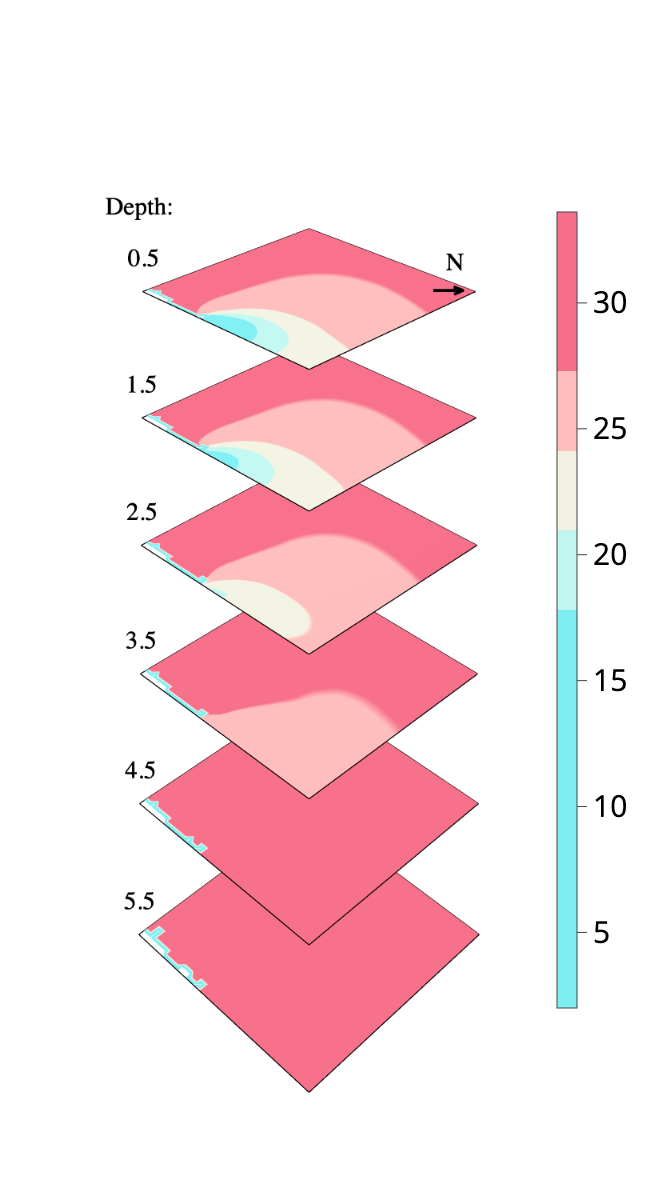}
         \caption{SINMOD prior}
         \label{fig:priorNA}
    \end{subfigure}
    \begin{subfigure}[b]{0.3\textwidth}
         \centering
         \includegraphics[width=\textwidth]{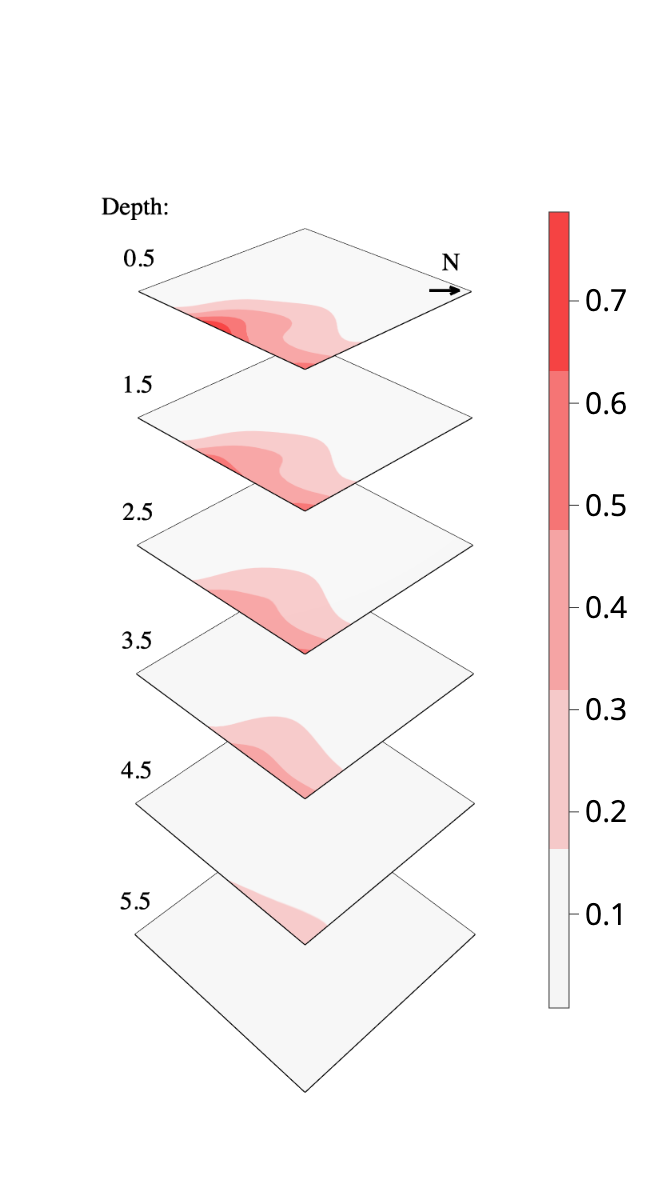}
         \caption{Marginal Variance}
         \label{fig:mvarapp}
    \end{subfigure}
    \begin{subfigure}[b]{0.3\textwidth}
         \centering
         \includegraphics[width=\textwidth]{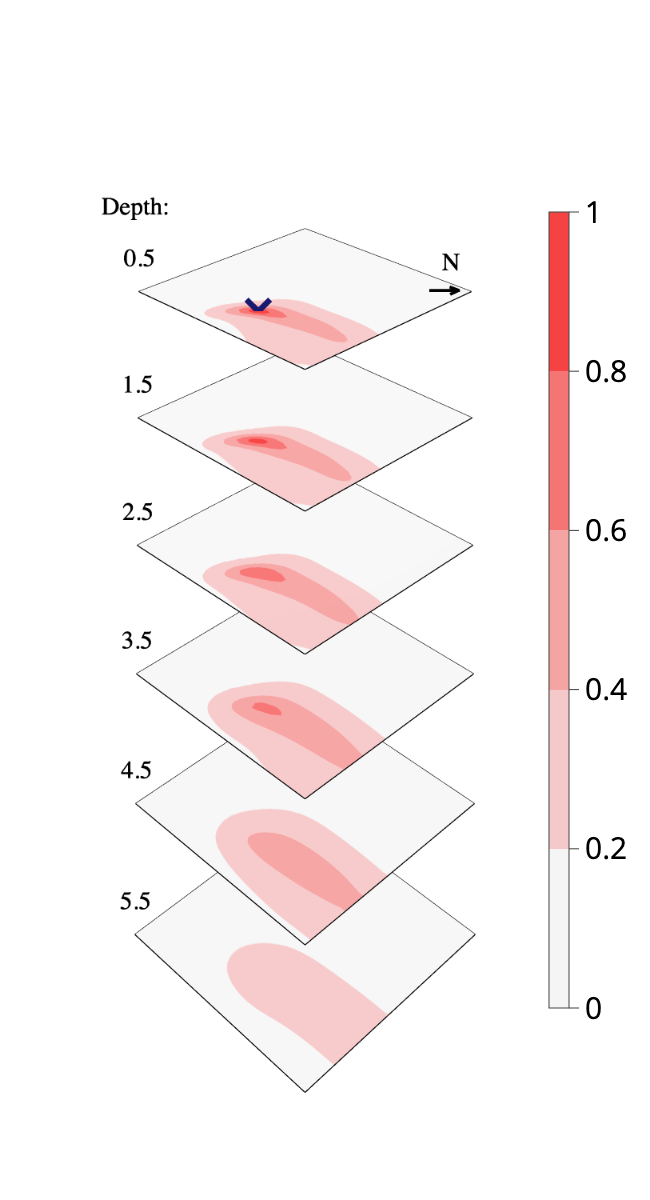}
         \caption{Correlation}
         \label{fig:corrapp}
    \end{subfigure}
    \caption{\label{fig:NAapp} Prior field \textbf{(a)} found from SINMOD simulations, the variance of the spatial effect \textbf{(b)} and spatial correlation of point [22,10,0] (marked) \textbf{(c)} in the non-stationary anisotropic model. The N-arrow is the cardinal north.}
\end{figure}

We fit the flexible non-stationary anisotropic model with 190 parameters, $\hat{\bm{\theta}}_{\mathrm{NA}}=(\bm{\alpha}_{\log\kappa},\bm{\alpha}_{\log\gamma},\bm{\alpha}_{v_x},\bm{\alpha}_{v_y},\bm{\alpha}_{v_z},\bm{\alpha}_{\rho_1},\bm{\alpha}_{\rho_2},\log \sigma_N^2)$, and the stationary anisotropic model with 8 parameters, $\hat{\bm{\theta}}_{\mathrm{SA}} = (\log \kappa^2, \log \gamma, v_x, v_y, v_z, \rho_1, \rho_2, \log \sigma_N^2)$, to the assumed independent realization from a GRF $\hat{\boldsymbol{\epsilon}}_1, \ldots.\hat{\boldsymbol{\epsilon}}_{143}$.
Note that there are $NMP = 22275$ spatial locations and the 144 empirical innovations cover the whole day of May 27, 2021.
Figures~\ref{fig:mvarapp} show the resulting variance of the spatial effect and Figure~\ref{fig:corrapp} the spatial correlation with location $(x,y,z) = (22,10,0)$ of the non-stationary anisotropic model. The same figures of the stationary anisotropic model can be found in Appendix~C, Figure~S3. 

In the next step, we construct the expected value of the GRF using the time average of the whole day, $\boldsymbol{\mu} = \sum_{t = 0}^{143}\boldsymbol{z}_t/144$. The mean is shown in Figure~\ref{fig:priorNA} and shows the overall tendency for freshwater near the river outlet and saltwater further out in the ocean. We choose the prior
\begin{equation}
    \boldsymbol{\eta} = \boldsymbol{\mu} + \boldsymbol{e},\label{eq:priorGRF}
\end{equation}
where we combine the fixed mean vector, $\boldsymbol{\mu}$, with a new realization, $\boldsymbol{e}$, of the estimated stationary anisotropic model or the non-stationary anisotropic model. This is a spatial prior on a $32\, \mathrm{m}\times 32 \, \mathrm{m}\times 1\, \mathrm{m}$ resolution.

\subsection{In-situ data collection and emulator evaluation}
\label{subsec:datacollect}

\begin{figure}[!htb]
    \centering
    \includegraphics[width=0.6\textwidth]{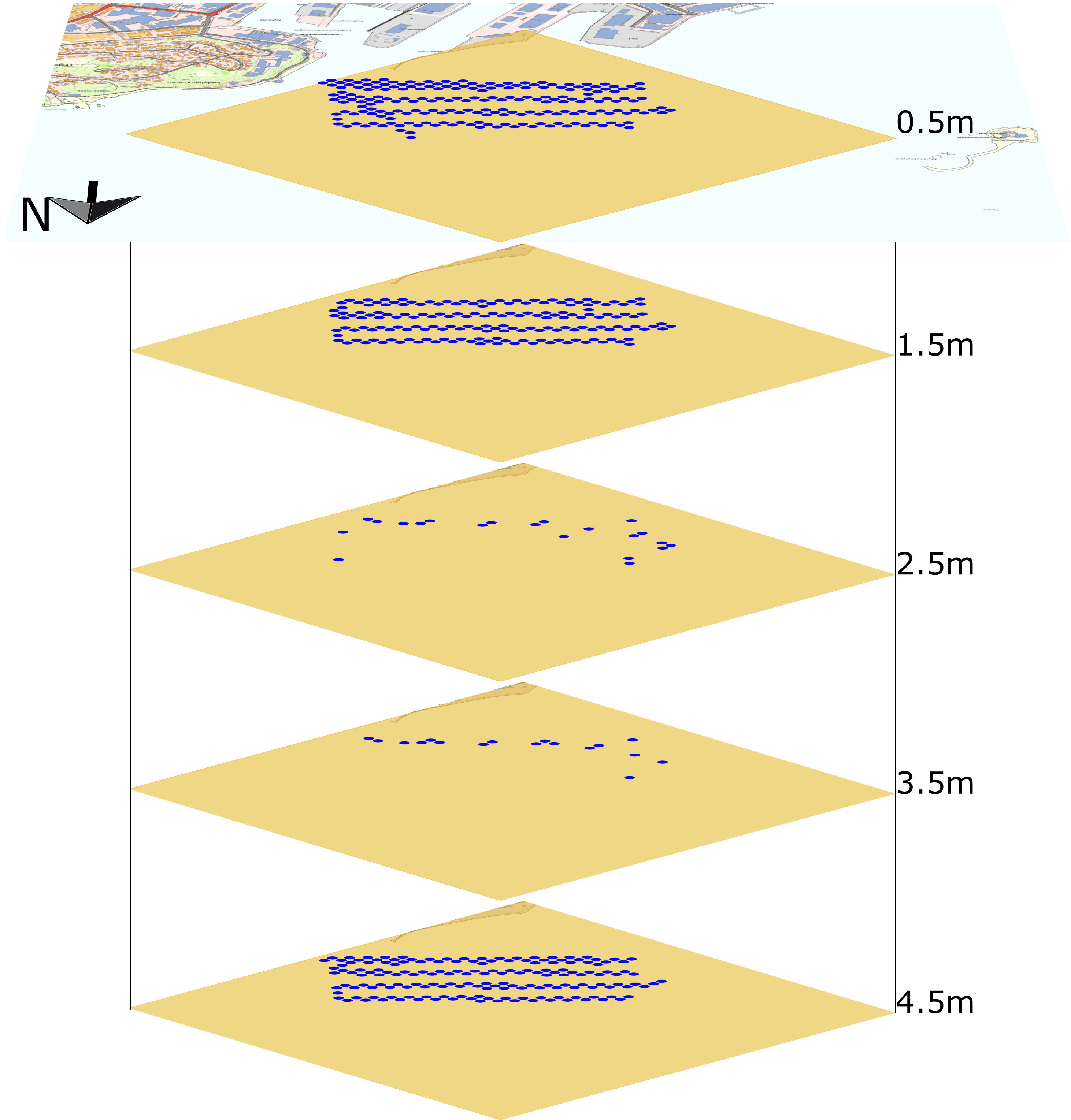}
    \caption{Measurement locations of the AUV in the top 6 depth layers of the spatial field on May 27th, 2021, in Trondheimsfjorden at Ladehammaren just outside of Trondheim, Norway. The N-arrow is the cardinal north.}
    \label{fig:datacollected}
\end{figure}

In-situ measurements were made with the AUV on May 27, 2021, between 10:30 and 14:30. The AUV followed 9 pre-planned paths within the area of operation: two intersects at 0.5m depth one northbound and one north-westbound starting from the river, two zig-zags in each depth layer (0.5m,2m,5m), and one up and down pattern in depth ranging from 0.5m to 10.5m moving north-westbound starting from the river. Figure~\ref{fig:datacollected} displays the locations of the measurements in the top 5 layers of the field. 

The AUV is moving at 1.5 m/s and continuously samples the salinity. This means that multiple measurements are made within each $32\, \mathrm{m}\times 32 \, \mathrm{m}\times 1\, \mathrm{m}$  grid cell. 
Measurements are represented as $y_i$, $i = 1, \ldots, n_{\mathrm{obs}}$, whereby $y_i$ is the average value measured in grid cell $i$. We combine these measurements with the prior in Equation \eqref{eq:priorGRF} using
\begin{align*}
    y_i|\boldsymbol{\eta}, \sigma_{\mathrm{N}}^2 &\overset{\text{ind}}{\sim} \mathcal{N}(\boldsymbol{a}_i^\mathrm{T}\boldsymbol{\eta}, \sigma_\mathrm{meas}^2), \quad i = 1, \ldots, n_{\mathrm{obs}}, \\
    \boldsymbol{\eta} &\sim \mathcal{N}(\boldsymbol{\mu}, \mathbf{Q}_{\mathrm{Prior}}^{-1}),
\end{align*}
where $\boldsymbol{a}_i$ selects the correct grid cell, $\mathbf{Q}_{\mathrm{Prior}}^{-1}$ is the estimated precision matrix for the GMRF, and the Gaussian likelihood with nugget variance $\sigma_\mathrm{meas}^2$ describes measurement noise and sub-grid variation. In general, we would estimate $\sigma_{\mathrm{meas}}^2$ using a trial run, but in this case, we estimated $\sigma_{\mathrm{meas}}^2$ using the average empirical variance over all observed grid cells in the total dataset. Note that we have not accounted for the uncertainty in the AUVs positions in these models. As the AUV dive, it loses its GPS signal and only relies on estimated location. When the GPS signal is returned a linear interpolation is made to account for drift but no uncertainty is included. 

\begin{figure}[!ht]
    \centering
    \includegraphics[width=0.8\textwidth]{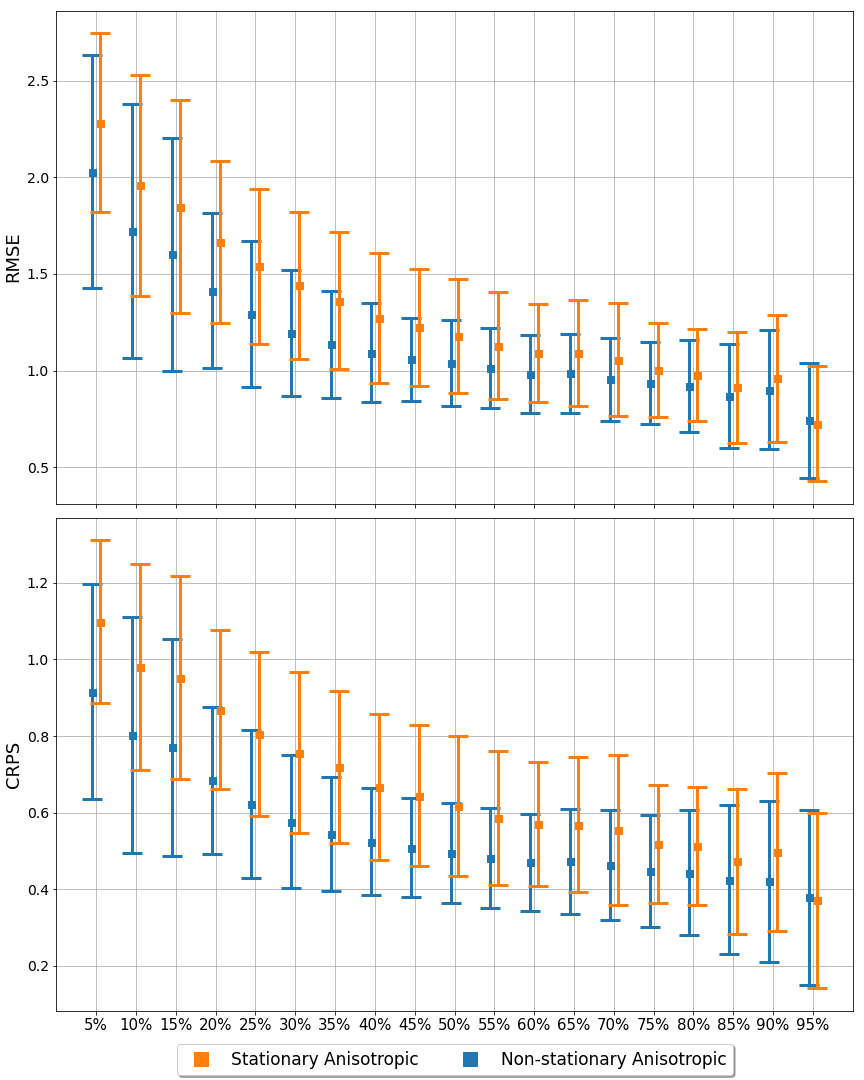}
    \caption{The root mean square error (RMSE, top) and the continuous ranked probability score (CRPS, bottom) of predictions from the stationary anisotropic (orange) and non-stationary anisotropic models (blue) given different proportions of observed data (5\%, 95\%). The error bars are the standard deviations of the different measures under random permutations of the 9 segments.}
    \label{fig:appres}
\end{figure}

We evaluated the two priors, or emulators, by randomly ordering the 9 segments and then sequentially including more and more observations for predicting the remaining hold-out data. The random permutation of the segments was done repeatedly to determine the variation in scores over different paths. This scheme evaluates the AUVs' ability to predict future observations while maintaining the sequential structure of measurements.
Figure \ref{fig:appres} shows that the non-stationary model provides a better prior for the salinity in the ocean than the stationary model. The differences are largest when little data is available, which is consistent with the idea that the prior is most important in this case. The non-stationary model can leverage knowledge about which areas are most uncertain using the spatially varying marginal variance and update the prior based on expected similarities from the spatially varying anisotropy. The improvements are seen both in point predictions through RMSE and in predictive distributions as measured by CPRS \citep{gneiting_strictly_2007}.

\section{Discussion}
We extend the class of SPDE-based GRFs introduced in \citet{fuglstad_exploring_2014} to three-dimensional space by overcoming two key issues: parametrization and computation. For the former, we developed a specification of spatially varying anisotropy through a spatially varying baseline isotropic dependence, and two orthogonal spatially varying vector fields that describe extra dependence. This allows for an interpretable description of the  $3\times 3$ positive definite matrix describing anisotropy. For the latter, we use a finite volume method to construct a GMRF that approximates the solution of the SPDE.

The specification of spatially varying marginal variance and spatially varying anisotropy requires specifying 7 spatially varying real functions. In this paper, we expand each function with a clamped B-spline basis. If each function uses $P^3$ basis functions, this gives in total $7P^3$ coefficients. As demonstrated in the simulation study, an unpenalized estimation of these parameters requires a densely observed area and multiple realizations. Application of the new models in data-sparse situations will require penalties that restrict the regularity of the 7 spatially varying functions. However, more research is needed to come up with a practical way to determine the appropriate strength of penalization for each of the functions.

While we did not experience any practical issues with the chosen way to describe the two orthogonal vector fields, the construction has a ``gimbal lock'' type issue. If one vector field points exactly along the $z$-axis, there is no unique choice for the second vector field. A potential way to avoid this issue is by describing the orientation of the two orthogonal vector fields through quaternions or Euler-Rodrigues parameters.

Moving from two-dimensional space to three-dimensional space introduces an asymptotically higher computation cost as a function of grid size. For a regular three-dimensional grid with $N$ nodes, the computational cost is $\mathcal{O}(N^2)$ compared to $\mathcal{O}(N^{3/2})$ in two-dimensional space. This increased computational cost arises from increased fill-in in the Cholesky factor. However, the application demonstrates that the use of a grid size of $N = 22275$ is unproblematic even for real-time updates on an AUV.

For the predictions of salinity in the Trondheim's fjord, we see the highest improvement of the complex GRF prior compared to an isotropic GRF, for sparse in-situ measurements. As more data is collected, the difference between the models decreases. This suggests that the key advantage of training the more complex GRF is to encode prior physical knowledge so that we can more effectively update knowledge about unobserved locations. Salinity was used as an example, but in general, the same approach could be used to map other biologically interesting quantities such as phytoplankton \citep{fossum2019toward}.
The GRFs developed in this paper are a step forward in quantifying beliefs about unobserved regions in the ocean, which is essential for optimal decisions and more effective autonomous sampling \citep{fossum2021learning}.

In future work, it would be interesting to add a dynamic component to the model to capture physical processes such as diffusion and advection. However, this substantially increases computational cost, and it is not clear to which degree an advection field from a numerical model should be trusted and which boundary conditions are best in an advection-dominated problem. The new class of GRFs shows great promise for encoding prior knowledge about a phenomenon in a computationally efficient way. However, overfitting is an important issue, and we must consider ways to penalize the complexity. In particular, we need to consider ways to allow flexibility in an area where it is needed such as a river outlet, and restrict flexibility in areas where we expect stationarity.

\section*{Acknowledgments}
  Berild and Fuglstad are supported by the Research Council of Norway, project number 305445. The authors are grateful to Ingrid Ellingsen and SINTEF for providing the simulations from the numerical ocean model SINMOD.

\newpage

\appendix
\setcounter{section}{0}%
\renewcommand{\thesection}{\Alph{section}.}
\setcounter{subsection}{0}%
\renewcommand{\thesubsection}{\Alph{section}.\arabic{subsection}}
\setcounter{table}{0}%
\renewcommand{\thetable}{S\arabic{table}}%
\setcounter{figure}{0}%
\renewcommand{\thefigure}{S\arabic{figure}}%
\setcounter{equation}{0}%
\renewcommand{\theequation}{S\arabic{equation}}%

\section{General properties}
\label{sec:genprop}

\subsection{Marginal Variance}
\label{subsec:mvar}
Here, we will derive the expression for the marginal variance in a general sense and then specify it for three-dimensional spaces with exponential covariance functions. The SPDE considered in this work is 
\begin{equation}
    (\kappa^2 - \nabla \cdot \matr{H}\nabla)^{\alpha/2}u(\bm{s}) = \mathcal{W}(\bm{s}), 
    \label{eq:SPDE}
\end{equation}
where $\bm{s} \in \mathcal{D} \subseteq \reals^d$ a spatial location in the domain of dimension $d$ and $\alpha = \nu + d/2$ where $\nu > 0$ is the smoothness. Any solution of this SPDE is a Matérn field and let $\sigma_m > 0$ be its marginal standard deviation; then, its covariance function is
\begin{equation}
    r(\bm{s}_1,\bm{s}_2) = \frac{\sigma_m^2}{2^{\nu - 1}\Gamma(\nu)}(\kappa||\matr{H}^{-1/2}(\bm{s}_1-\bm{s}_2)||)^\nu K_\nu (\kappa||\matr{H}^{-1/2}(\bm{s}_1-\bm{s}_2)||).
    \label{eq:expcovfunc}
\end{equation}

The transfer function of the SPDE is
\begin{equation*}
    g(\bm{w}) = (\kappa^2 + \bm{w}^T\matr{H}\bm{w})^{-\alpha/2}.
\end{equation*}
Using this and by including the spectral density of standard Gaussian white noise in $\reals^d$ is $(2\pi)^{-d}$, 
the spectral density of the solution of the SPDE is 
\begin{equation*}
    f_S(\bm{w}) = (2\pi)^{-d} (\kappa^2 + \bm{w}^T\matr{H}\bm{w})^{-\alpha}.
\end{equation*}
Lastly, to find the marginal variance of the field the integral of the spectral density is made over $\reals^d$ as
\begin{equation*}
    \sigma_m^2 = \int_{\reals^d} f_S(\bm{w})\dd\bm{w}.
\end{equation*}
Including the change of variables $\bm{w} = \kappa \matr{H}^{-1/2}\bm{z}$ the expression becomes
\begin{equation}
    \begin{aligned}
    \sigma_m^2  & = (2\pi)^{-d} \int_{\reals^d} (\kappa^2 + \kappa^2\bm{z}^T\bm{z})^{-\alpha} \det(\kappa\matr{H}^{-1/2}) \dd\bm{z}\\
     & = (2\pi)^{-d}\int_{\reals^d}\kappa^{d-2\alpha}(1+\bm{z}^T\bm{z})^{-\alpha} \det(\matr{H})^{-1/2} \dd\bm{z} \\
     & \overset{\alpha = \nu + d/2}{=} (2\pi)^{-d}\kappa^{-2\nu}\det(\matr{H})^{-1/2}\int_{\reals^d}(1+\bm{z}^T\bm{z})^{-\alpha} \dd\bm{z},
    \end{aligned}
    \label{eq:derivmvar}
\end{equation}
which by specifying a exponential covariance in $\reals^3$ with $\alpha = 2$, $\nu = 1/2$ and $d=3$ is 
\begin{equation*}
    \sigma_m^2  = \frac{1}{8\pi \kappa \sqrt{\det(\matr{H})}}.
\end{equation*}
Note that the integral in Equation~\eqref{eq:derivmvar} is solved by converting to polar coordinates as
\begin{equation*}
        \int_{\reals^3}\frac{1}{(1+\bm{z}^T\bm{z})^2} \dd\bm{z} = \int_0^\pi \sin(\phi)\dd\phi \int_0^{2\pi}\dd\theta \int_0^\infty \frac{\rho^2}{(1+\rho^2)^2}\dd\rho = \pi^2.
\end{equation*}

\subsection{Covariance function}
\label{subsec:covfunc}

Evaluating Equation~\eqref{eq:expcovfunc} at $\nu = 1/2$ and including the expression for the marginal variance the covariance function can be formalized as
\begin{equation*}
    r(\bm{s}_1,\bm{s}_2) = \sqrt{\frac{2}{\pi}}\frac{1}{8\pi\kappa\sqrt{\det(\matr{H})}}\sqrt{\kappa||\matr{H}^{-1/2}(\bm{s}_1-\bm{s}_2)||} K_{\frac{1}{2}} (\kappa||\matr{H}^{-1/2}(\bm{s}_1-\bm{s}_2)||).
\end{equation*}
Then, consider the modified Bessel function of the second kind
\begin{equation*}
    K_{n}(z) = \sqrt{\frac{\pi}{2z}} \frac{e^{-z}}{(n-\frac{1}{2})!} \int_0^\infty e^{-t}t^{n-1/2} \left(1 - \frac{t}{2z}\right)^{n-1/2}\dd t,
\end{equation*}
and evaluate this at order 1/2 gives
\begin{equation*}
    K_{\frac{1}{2}}(z) = \sqrt{\frac{\pi}{2z}} e^{-z}.
\end{equation*}
The covariance function can then be formalized as
\begin{equation}
    \begin{aligned}
    r\left(\bm{s}_1,\bm{s}_2\right) = & \sqrt{\frac{2}{\pi}}\sigma_m^2\sqrt{\kappa||\matr{H}^{-1/2}(\bm{s}_1-\bm{s}_2)||} \\ 
     & \times \sqrt{\frac{\pi}{2\cdot \kappa||\matr{H}^{-1/2}(\bm{s}_1-\bm{s}_2)||}}\exp\left(-\kappa||\matr{H}^{-1/2}(\bm{s}_1-\bm{s}_2)||\right)\\
     = & \sigma_m^2\exp\left(-\kappa||\matr{H}^{-1/2}(\bm{s}_1-\bm{s}_2)||\right).
    \end{aligned}
    \label{eq:expcovfunc3}
\end{equation}

\subsection{One-dimensional clamped B-splines}
\label{app:bSpline1D}
We illustrate the construction of 1-dimensional splines B-splines using the interval $[A,B]\in \reals$. Let $A = t_0 < t_1 < \cdots < t_m = B$ be the knot points. Then the zero-order B-splines are constructed recursively as
\[
    B_{i,0}(t) = \begin{cases} 1, & t_i \leq t \leq t_{i+1}, \\ 0, & \text{otherwise}, \end{cases}, \quad t\in[A,B],
\]
for $i = 0, \ldots, p-1$.
Let $r$ denote the order of the B-splines. 
The first- and second-order basis splines are constructed as
\begin{equation*}
B_{i,r}(t) = \frac{t-t_i}{t_{i+r}-t_i}B_{i,r-1}(t) + \frac{t_{i+r+1}-t}{t_{i+r+1}-t_{i+1}}B_{i+1,r-1}(t), \quad t\in[A,B],
\end{equation*}
for $i = 0, \ldots, p-r-1$.

Using the $r$-order B-spline basis, we construct a function $g:[A,B]\rightarrow \reals$ by
\[
    g(t) = \sum_{i = 0}^{p-r-1} \alpha_i B_{i,r}(t).
\]
where $\alpha_0, \ldots, \alpha_{p-r-1}\in \reals$ are coefficients.
We use a clamped spline where $g'(A) = g'(B) = 0$ and need the additional requirement that $\alpha_0 = \alpha_1$ and $\alpha_{p-r-2} = \alpha_{p-r-1}$.

\subsection{Integrated likelihood}
\label{subsec:likelihood}

The distribution of $\bm{z} = (\bm{u},\bm{\beta})$ is given by
\begin{equation*}
    \bm{z}|\bm{\theta} \sim \norm(\bm{0},\matr{Q}_z^{-1}),
\end{equation*}
and the observation model is 
\begin{equation*}
    \bm{y}|\bm{z},\bm{\theta},\sigma_N^2 \sim \norm_n(\matr{S}\bm{z},\matr{I}_n\sigma_N^2).
\end{equation*}
From this the distribution of $\bm{z}$ given some observations $\bm{y}$ is
\begin{equation*}
    \begin{aligned}
    \pi(\bm{z}|\bm{\theta},\sigma_N^2,\bm{y}) & \propto \pi(\bm{z},\bm{\theta},\sigma_N^2,\bm{y}) \\
    & = \pi(\bm{\theta},\sigma_N^2)\pi(\bm{z}|\bm{\theta})\pi(\bm{y}|\bm{\theta},\sigma_N^2,\bm{z}) \\
    & \propto \exp\left(-\frac{1}{2}\bm{z}^\mathrm{T}\matr{Q}_z\bm{z} - \frac{1}{2}(\bm{y}-\matr{S}\bm{z})^\mathrm{T}\matr{I}_n \sigma_N^{-2} (\bm{y}-\matr{S}\bm{z})\right) \\
    & \propto \exp\left(-\frac{1}{2}\left(\bm{z}^\mathrm{T}\left(\matr{Q}_z + \sigma_N^{-2} \matr{S}^\mathrm{T}\matr{S}\right)\bm{z} - 2\bm{z}^\mathrm{T}\matr{S}^T\bm{y}\cdot \sigma_N^{-2}\right)\right)\\
    & \propto \exp\left(-\frac{1}{2}(\bm{z}-\bm{\mu}_C)^\mathrm{T}\matr{Q}_C(\bm{z}-\bm{\mu}_C)\right)\\
    & \Downarrow\\
    \bm{z}|\bm{\theta},\sigma_N^2,\bm{y} &\sim \norm_n\left(\bm{\mu}_C , \matr{Q}_C^{-1}\right)
    \end{aligned}
\end{equation*}
Here, $\matr{Q}_C = \matr{Q}_z + \matr{S}^\mathrm{T}\matr{S}\cdot\sigma_N^{-2}$ is the conditional precision matrix and $\bm{\mu_C} = \matr{Q}_C^{-1}\matr{S}^\mathrm{T}\bm{y}\cdot\sigma_N^{-2}$ is the conditional mean. 

Then, integrating out $\bm{z}$ from the joint distribution gives
\begin{equation*}
    \begin{aligned}
    \pi(\bm{\theta},\sigma_N^2,\bm{y}) & = \frac{\pi(\bm{\theta},\bm{z},\sigma_N^2,\bm{y})}{\pi(\bm{z}|\bm{\theta},\sigma_N^2,\bm{y})} \\
     & = \frac{\pi(\bm{\theta},\sigma_N^2)\pi(\bm{z}|\bm{\theta})\pi(\bm{y}|\bm{\theta},\sigma_N^2,\bm{z})}{\pi(\bm{z}|\bm{\theta},\sigma_N^2,\bm{y})},
    \end{aligned}
\end{equation*}
where the left-hand side does not depend on $\bm{z}$ such that it may be evaluated for any given value. Let us evaluate it for $\bm{z}= \bm{\mu}_C$ such that
\begin{equation*}
    \begin{aligned}
    \pi(\bm{\theta},\sigma_N^2,\bm{y}) \propto & \frac{\pi(\bm{\theta},\sigma_N^2)\pi(\bm{z} = \bm{\mu}_C|\bm{\theta})\pi(\bm{y}|\bm{\theta},\sigma_N^2,\bm{z}=\bm{\mu}_C)}{\pi(\bm{z}=\bm{\mu}_C|\bm{\theta},\sigma_N^2,\bm{y})} \\
    \propto & \pi(\bm{\theta})\frac{|\matr{Q}_z|^{1/2}|\matr{I}_n\cdot\sigma_N^{-2}|^{1/2}}{|\matr{Q}_C|^{1/2}}\exp\left(-\frac{1}{2}\bm{\mu}_C^\mathrm{T}\matr{Q}_z\bm{\mu}_C\right)\\
    & \times \exp\left(-\frac{1}{2}(\bm{y}-\matr{S}\bm{\mu}_C)^\mathrm{T}\matr{I}_n\cdot \sigma_N^{-2} (\bm{y}-\matr{S}\bm{\mu}_C) \right).
    \end{aligned}
\end{equation*}
The last term $\pi(\bm{z}|\bm{\theta},\sigma_N^2,\bm{y})$ is removed since it is equal to 1. Thereby, conditioning on $\bm{y}$ and taking the log we have the log-likelihood
\begin{equation}
\begin{aligned}
    \log(\pi(\bm{\theta},\sigma_N^2|\bm{y})) = & \mathrm{Constant} + \log(\pi(\bm{\theta},\sigma_N^2)) + \frac{1}{2}\log(\det(\matr{Q}_z)) + \frac{n}{2}\log(\sigma_N^{-2}) \\
     & - \frac{1}{2}\log(\det(\matr{Q}_C)) - \frac{1}{2}\bm{\mu}_C^\mathrm{T}\matr{Q}_z\bm{\mu}_C - \frac{1}{2\cdot\sigma_N^2}(\bm{y}-\matr{S}\bm{\mu}_C)^\mathrm{T}(\bm{y}-\matr{S}\bm{\mu}_C).
\end{aligned}
\label{eq:loglike}
\end{equation}

\subsection{Gradient of the log-likelihood}
\label{subsec:gradient}
This section is similar to the derivation of the gradient presented in the supplementary material of \citet{fuglstad_does_2015}.
\begin{equation*}
\begin{aligned}
    \log(\pi(\bm{\theta},\tau_N|\bm{y})) = & \mathrm{Constant} + \log(\pi(\bm{\theta},\tau_N)) + \frac{1}{2}\log(\det(\matr{Q}_z)) + \frac{n}{2}\log(\sigma_N^{-2}) \\
     & - \frac{1}{2}\log(\det(\matr{Q}_C)) + \frac{1}{2}\bm{\mu}_C^\mathrm{T}\matr{Q}_C\bm{\mu}_C - \frac{\tau_N}{2}\bm{y}^\mathrm{T}\bm{y}.
\end{aligned}
\end{equation*}
Note that the last two terms are rewritten for simplicity in the gradient calculation and that the variance of the Gaussian noise term, $\sigma_N^2$ is re-parametrized with its inverse $\tau_N = 1/\sigma_N^2$ (precision). Derivatives of the log-likelihood are taken with respect to $\theta_i$, the elements of $\bm{\theta}$, and the precision on log scale as $\log(\tau_N)$.

The first term is a constant and therefore its derivative is zero with respect to any of the parameters. The next term, the penalty or the prior of the parameters, is not used in this paper and otherwise depends on the choice of penalty so gradient calculation is not specified for this term. 

To continue note the derivatives of the precision matrix
\begin{equation*}
    \frac{\partial \matr{Q}_C }{\partial \theta_i} = \frac{\partial \matr{Q}_z }{\partial \theta_i} \enspace\enspace \mathrm{and} \enspace\enspace
    \frac{\partial \matr{Q}_C }{\partial \log(\tau_N)} = \matr{S}^T\matr{S}\tau_N,
\end{equation*}
which is used in the following derivations. First, the derivatives with respect to $\theta_i$ are considered. The derivative of the log determinant terms are 
\begin{equation*}
\begin{aligned}
\frac{\partial}{\partial \theta_i}\left(\log(\det(\matr{Q})) - \log(\det(\matr{Q}_C))\right) = & \mathrm{Tr}\left(\matr{Q}^{-1}\frac{\partial \matr{Q}}{\partial \theta_i}\right) -\mathrm{Tr}\left(\matr{Q}_C^{-1}\frac{\partial \matr{Q}}{\partial \theta_i}\right) \\
= & \mathrm{Tr}\left((\matr{Q}^{-1} - \matr{Q}_C^{-1})\frac{\partial \matr{Q}}{\partial \theta_i}\right),
\end{aligned}
\end{equation*}
and the derivative of the quadratic terms are 
\begin{equation*}
\begin{aligned}
\frac{\partial}{\partial \theta_i}\left(\frac{1}{2}\bm{y}^\mathrm{T}\bm{y}\tau_N + \frac{1}{2}\bm{\mu}_C^\mathrm{T}\matr{Q}_C\bm{\mu}_C\right) = & \frac{\partial}{\partial \theta_i}\left(\frac{1}{2}\bm{\mu}_C^\mathrm{T}\matr{Q}_C\bm{\mu}_C\right) \\
= & -\frac{1}{2}\bm{y}^\mathrm{T}\tau_N \matr{S}\matr{Q}_C^{-1}\left(\frac{\partial \matr{Q}_C}{\partial \theta_i}\right)\matr{Q}_C^{-1}\matr{S}^\mathrm{T}\tau_N \bm{y}\\
= & -\frac{1}{2}\bm{\mu}_C^\mathrm{T}\left(\frac{\partial \matr{Q}}{\partial \theta_i} \right)\bm{\mu}_C.
\end{aligned}
\end{equation*}
Then, combining these the derivative of the log-likelihood with respect to $\theta_i$ is 
\begin{equation*}
\frac{\partial}{\partial \theta_i}\log(\pi(\bm{\theta},\tau_N|\bm{y})) =  \frac{\partial}{\partial \theta_i}\log(\pi(\bm{\theta},\tau_N)) + \mathrm{Tr}\left((\matr{Q}^{-1} - \matr{Q}_C^{-1})\frac{\partial \matr{Q}}{\partial \theta_i}\right) - \frac{1}{2}\bm{\mu}_C^\mathrm{T}\left(\frac{\partial \matr{Q}}{\partial \theta_i} \right)\bm{\mu}_C
\end{equation*}

Next, the derivative with respect to the log precision, $\log \tau_N$, is considered. The derivative of the log determinant terms are
\begin{equation*}
\begin{aligned}
    \frac{\partial}{\partial \log(\tau_N)}\left(\frac{n}{2} \log(\tau_N) - \frac{1}{2}\log(\det(\matr{Q}_C))\right) = & \frac{n}{2}- \frac{1}{2}\mathrm{Tr}\left(\matr{Q}_C^{-1}\frac{\partial}{\partial \log(\tau_N)}\matr{Q}_C\right) \\
    = & \frac{n}{2}-\frac{1}{2}\mathrm{Tr}\left(\matr{Q}_C^{-1}\matr{S}^\mathrm{T}\matr{S}\cdot \tau_N\right)
\end{aligned}
\end{equation*}
Further, the derivative of $1/2\bm{y}^\mathrm{T}\bm{y}\cdot\tau_N$ with respect to $\log(\tau_N)$ is just the same expression so the remaining quadratic term becomes
\begin{equation*}
    \begin{aligned}
        \frac{\partial\frac{1}{2}\bm{\mu}_C^\mathrm{T}\matr{Q}_C\bm{\mu}_C}{\partial \log(\tau_N)} = & \frac{\partial\frac{1}{2}\bm{y}^\mathrm{T}\tau_N\matr{S}\matr{Q}_C^{-1}\matr{S}^\mathrm{T}\tau_N\bm{y}}{\partial \log(\tau_N)} \\
        = & \bm{y}^\mathrm{T}\tau_N\matr{S}\matr{Q}_C^{-1}\matr{S}^\mathrm{T}\frac{\partial \tau_N}{\partial \log(\tau_N)}\bm{y} - \frac{1}{2}\bm{y}^\mathrm{T}\tau_N\matr{S}\matr{Q}_C^{-1} \frac{\partial \matr{Q}_C}{\partial \log(\tau_N)}\matr{Q}_C^{-1}\matr{S}^\mathrm{T}\tau_N\bm{y} \\
        = & \bm{\mu}_C^\mathrm{T}\matr{S}^\mathrm{T}\tau_N\bm{y} - \frac{1}{2}\bm{\mu}_C^\mathrm{T}\matr{S}^\mathrm{T}\matr{S}\bm{\mu}_C\tau_N,
    \end{aligned}
\end{equation*}
and then, by adding the last quadratic term, the expression simplifies to
\begin{equation*}
    - 1/2\bm{y}^\mathrm{T}\bm{y}\cdot\tau_N + \bm{\mu}_C^\mathrm{T}\matr{S}^\mathrm{T}\bm{y}\cdot\tau_N - \frac{1}{2}\bm{\mu}_C^\mathrm{T}\matr{S}^\mathrm{T}\matr{S}\bm{\mu}_C\cdot\tau_N =  -\frac{1}{2} (\bm{y} - \matr{S}\bm{\mu}_C)^\mathrm{T}(\bm{y}- \matr{S}\bm{\mu}_C)\cdot \tau_N.
\end{equation*}
Finally, combining all these terms we have the derivative of the log-likelihood with respect to $\log(\tau_N)$:
\begin{equation*}
    \begin{aligned}
        \frac{\partial \log(\pi(\bm{\theta},\tau_N|\bm{y}))}{\partial \log(\tau_N))} = & \frac{\partial \log(\pi(\bm{\theta},\tau_N)}{\partial \log(\tau_N)} + \frac{n}{2}-\frac{1}{2}\mathrm{Tr}\left(\matr{Q}_C^{-1}\matr{S}^\mathrm{T}\matr{S}\cdot \tau_N\right)\\
        & -\frac{1}{2} (\bm{y} - \matr{S}\bm{\mu}_C)^\mathrm{T}(\bm{y}- \matr{S}\bm{\mu}_C)\cdot \tau_N
    \end{aligned}
\end{equation*}

Note that the derivative of $\matr{Q}_C$ can be calculated quickly and it is derived from a series of chain rules; first on $\matr{Q}_C$, then on $\matr{A}$ and $\matr{A}_\matr{H}$, and finally within $\matr{H}$. The most computationally heavy calculation in the gradient of the log-likelihood is to calculate the inverses in the difference $\matr{Q}^{-1} - \matr{Q}_C^{-1}$. However, since this term is multiplied with the derivative of $\matr{Q}$ with respect to $\theta_i$, which carries the non-zero structure of $\matr{Q}$, only elements of $\matr{Q}^{-1}$ and $\matr{Q}^{-1}_C$ which correspond to the non-zero structure of $\matr{Q}$ need to be calculated. This is done by calculating a partial inverse of two matrices as described in \citet{rue_markov_2010}.

\section{Derivation}
\label{sec:derivation}

\subsection{Discretization}
\label{subsec:disc}

\begin{figure}[!ht]
    \centering
    \includegraphics[width=0.9\textwidth]{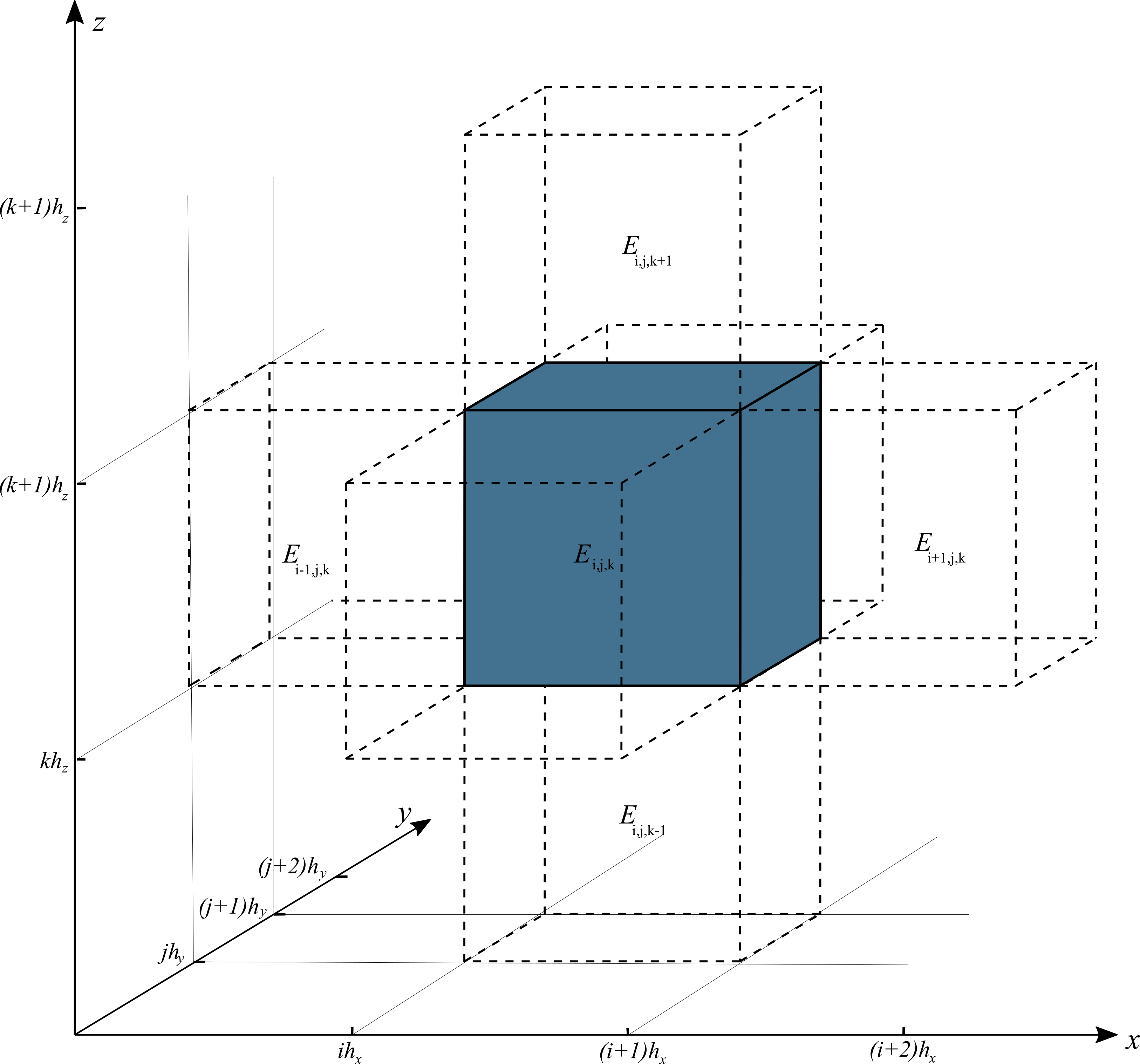}
    \caption{One cell $E_{i,j,k}$ in the discretization with its closest neighbours; $E_{i+1,j,k}$, $E_{i-1,j,k}$, $E_{i,j+1,k}$, $E_{i,j-1,k}$, $E_{i,j,k+1}$, and $E_{i,j,k-1}$.}
    \label{fig:reggrid}
\end{figure}

To find the local solution of the SPDE the domain $\mathcal{D} = [A_1, B_1]\times[A_2, B_2]\times[A_3, B_3]$ is divided into equally sized rectangular cubes or cells. We use $M$ cells to divide $[A_1, B_1]$ in the $x$-direction, $N$ cells on $[A_2, B_1]$ in $y$-direction and $P$ cells on $[A_{3}, B_{3}]$ in $z$-direction. The cells have sides parallel to each axis of size $h_x = (B_1-A_1)/M$, $h_y = (B_2-A_2)/N$, and $h_z = (B_3-A_3)/P$. The cells are assigned an index with regards to their cell number along each axes starting from number 0; $i\in[0, M]$ along $x$, $j\in[0, N]$ along $y$, and $k\in[0, P]$ along $z$. For a specific cell, its domain can be denoted as
\begin{equation*}
    E_{i,j,k} = [ih_x, (i+1)h_x]\times[jh_y,(j+1)h_y]\times[kh_z,(k+1)h_z],
\end{equation*}
and Figure~\ref{fig:reggrid} shows this cell and its closest neighbors. 
Furthermore,  as a regular grid is employed the volume of a cell is $V = h_x h_y h_z$.

\begin{figure}[!ht]
    \centering
    \includegraphics[width=0.9\textwidth]{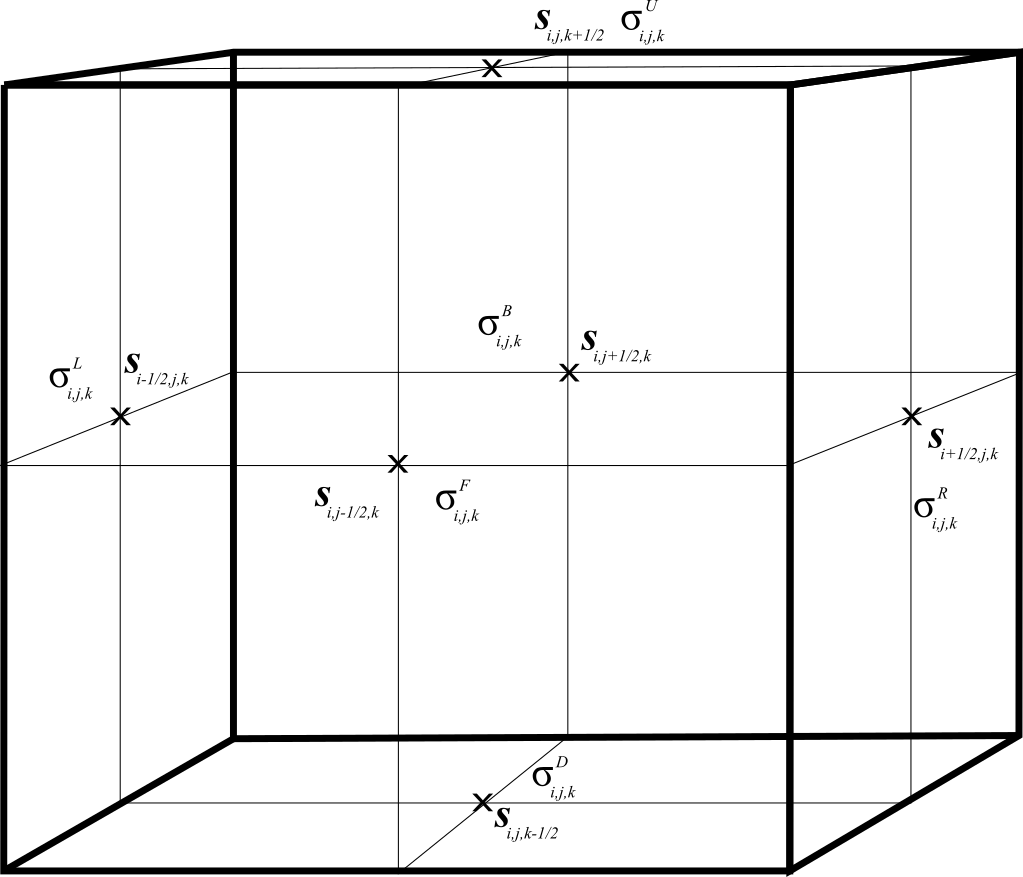}
    \caption{One cell $E_{i,j,k}$ of the discretization with all its faces; $\sigma_{i,j,k}^F$ (front), $\sigma_{i,j,k}^B$ (back), $\sigma_{i,j,k}^L$ (left), $\sigma_{i,j,k}^R$ (right), $\sigma_{i,j,k}^U$ (up), and $\sigma_{i,j,k}^D$ (down) each with its respective face centres.}
    \label{fig:disc}
\end{figure}

To further define the local solution of the SPDE we denote the faces of a grid cell as $\sigma_{i,j,k}^F$ (front), $\sigma_{i,j,k}^B$ (back), $\sigma_{i,j,k}^L$ (left), $\sigma_{i,j,k}^R$ (right), $\sigma_{i,j,k}^U$ (up) and $\sigma_{i,j,k}^D$ (down) with their respective face centers $\bm{s}_{i,j-1/2,k}$, $\bm{s}_{i,j+1/2,k}$, $\bm{s}_{i-1/2,j,k}$, $\bm{s}_{i+1/2,j,k}$, $\bm{s}_{i,j,k+1/2}$ and $\bm{s}_{i,j,k-1/2}$. Figure~\ref{fig:disc} describes the different faces of a cell. 

\subsection{Local solution of the SPDE}
\label{subsec:locsol} 

Note that this description is an extension to three dimensions of the derivation described in \citet{fuglstad_exploring_2014}, and the reader is referred to there for further details. 
To locally solve the SPDE a finite volume scheme is derived. First, Equation~\eqref{eq:SPDE} is integrated over a cell $E_{i,j,k}$ as 
\begin{equation}
    \int_{E_{ijk}} \kappa^2(\bm{s})u(\bm{s})\mathrm{d}\bm{s} - \int_{E_{ijk}} \nabla\cdot\matr{H}(\bm{s})\nabla u(\bm{s})\mathrm{d}\bm{s} = \int_{E_{ijk}} \mathcal{W}(\bm{s}) \mathrm{d}\bm{s},
    \label{eq:SPDEint1}
\end{equation}
where $\dd\bm{s}$ is a volume element. The integral of the Gaussian white noise on the right-hand side is a Gaussian variable with mean zero and variance equal to the volume of a cell which is independent of neighboring cells. Let $z_{ijk}$ be an standard Gaussian variable; then, Equation~\eqref{eq:SPDEint1} becomes
\begin{equation*}
    \int_{E_{ijk}} \kappa^2(\bm{s})u(\bm{s})\mathrm{d}\bm{s} - \int_{E_{ijk}} \nabla\cdot\matr{H}(\bm{s})\nabla u(\bm{s})\mathrm{d}\bm{s} = \sqrt{V} z_{ijk}.
\end{equation*}
Then, applying the divergence theorem to the second integral with the divergence operator gives
\begin{equation*}
    \int_{E_{ijk}} \kappa^2(\bm{s})u(\bm{s})\mathrm{d}\bm{s} - \oint_{\partial E_{ijk}} (\matr{H}(\bm{s})\nabla u(\bm{s}))^T \bm{n}(\bm{s}) \mathrm{d}\sigma = \sqrt{V} z_{ijk}.
\end{equation*}
The first integral is approximated by letting $k_{ijk}^2$ be the average value of the continuous function $\kappa^2(\bm{s})$ within a cell, i.e. $\kappa_{ijk}^2 = 1/V \int_{E_{ijk}} \kappa^2(\bm{s})\dd \bm{s}$, resulting in
\begin{equation}
    V \kappa^2_{ijk}u_{ijk} - \oint_{\partial E_{ijk}} (\matr{H}(\bm{s})\nabla u(\bm{s}))^T \bm{n}(\bm{s}) \mathrm{d}\sigma = \sqrt{V} z_{ijk}.
    \label{eq:SPDEint2}
\end{equation}

To describe the solution of the second integral it is divided into integrals over each surface as
\begin{equation}
     \oint_{\partial E_{ijk}} (\matr{H}(\bm{s})\nabla u(\bm{s}))^T \bm{n}(\bm{s}) \mathrm{d}\sigma  = W^L_{ijk} + W^R_{ijk} + W^B_{ijk} + W^F_{ijk} + W^U_{ijk} + W^D_{ijk},
\end{equation}
or $W^{\mathrm{dir}}_{ijk} =  \int_{\sigma^{\mathrm{dir}}_{ijk}} (\matr{H}(\bm{s})\nabla u(\bm{s}))^T \bm{n}(\bm{s}) \mathrm{d}\sigma$, where $\mathrm{dir}$ denotes the surface; $R$ (positive x-direction), $L$ (negative x-direction), $B$ (positive y-direction), $F$ (negative y-direction), $U$ (positive z-direction), and $D$ (negative z-direction). 
Now, an approximation of this surface integral over each face is required. It is assumed that the gradient of $u(\bm{s})$ is constant over each face and equal to the value at the center of each face.
The resulting scheme for the gradient on each face is described in Table~\ref{tab:scheme}.
\setlength\tabcolsep{2.0pt} 
\begin{table}[!htb]
    \centering
    \begin{tabular}{|c|c|}\hline
         Face & Scheme\\\hline
         $\sigma_{i,j,k}^R$ & \begin{tabular}{l}
                                $\frac{\partial}{\partial x} u(\bm{s}_{i+1/2,j,k}) \simeq \frac{1}{h_x}\left(u(\bm{s}_{i+1,j,k}) - u(\bm{s}_{i,j,k}) \right)$ \\
                                $\frac{\partial}{\partial y} u(\bm{s}_{i+1/2,j,k})  \simeq \frac{1}{4h_y}\left(u(\bm{s}_{i+1,j+1,k}) + u(\bm{s}_{i,j+1,k}) - u(\bm{s}_{i+1,j-1,k}) - u(\bm{s}_{i,j-1,k}) \right)$\\
                                $\frac{\partial}{\partial z} u(\bm{s}_{i+1/2,j,k})  \simeq \frac{1}{4h_z}\left(u(\bm{s}_{i+1,j,k+1}) + u(\bm{s}_{i,j,k+1}) - u(\bm{s}_{i+1,j,k-1}) - u(\bm{s}_{i,j,k-1}) \right)$
                             \end{tabular}\\\hline
        $\sigma_{i,j,k}^L$ & \begin{tabular}{l}
                             $\frac{\partial}{\partial x} u(\bm{s}_{i-1/2,j,k}) \simeq \frac{1}{h_x}\left(u(\bm{s}_{i,j,k}) - u(\bm{s}_{i-1,j,k}) \right)$\\
                             $\frac{\partial}{\partial y} u(\bm{s}_{i-1/2,j,k})  \simeq \frac{1}{4h_y}\left(u(\bm{s}_{i,j+1,k}) + u(\bm{s}_{i-1,j+1,k}) - u(\bm{s}_{i,j-1,k}) - u(\bm{s}_{i-1,j-1,k}) \right)$\\
                             $\frac{\partial}{\partial z} u(\bm{s}_{i-1/2,j,k})  \simeq \frac{1}{4h_z}\left(u(\bm{s}_{i,j,k+1}) + u(\bm{s}_{i-1,j,k+1}) - u(\bm{s}_{i,j,k-1}) - u(\bm{s}_{i-1,j,k-1}) \right)$
                             \end{tabular}\\\hline
        $\sigma_{i,j,k}^B$ & \begin{tabular}{l}
                             $\frac{\partial}{\partial x} u(\bm{s}_{i,j+1/2,k}) \simeq\frac{1}{4h_x}\left(u(\bm{s}_{i+1,j+1,k}) + u(\bm{s}_{i+1,j,k}) - u(\bm{s}_{i-1,j+1,k}) - u(\bm{s}_{i-1,j,k}) \right)$\\
                             $\frac{\partial}{\partial y} u(\bm{s}_{i,j+1/2,k})  \simeq \frac{1}{h_y}\left(u(\bm{s}_{i,j+1,k}) - u(\bm{s}_{i,j,k}) \right)$\\
                             $\frac{\partial}{\partial z} u(\bm{s}_{i,j+1/2,k})  \simeq \frac{1}{4h_z}\left(u(\bm{s}_{i,j+1,k+1}) + u(\bm{s}_{i,j,k+1}) - u(\bm{s}_{i,j+1,k-1}) - u(\bm{s}_{i,j,k-1}) \right)$
                             \end{tabular}\\\hline       
        $\sigma_{i,j,k}^F$ & \begin{tabular}{l}
                             $\frac{\partial}{\partial x} u(\bm{s}_{i,j-1/2,k}) \simeq\frac{1}{4h_x}\left(u(\bm{s}_{i+1,j,k}) + u(\bm{s}_{i+1,j-1,k}) - u(\bm{s}_{i-1,j,k}) - u(\bm{s}_{i-1,j-1,k}) \right)$\\
                             $\frac{\partial}{\partial y} u(\bm{s}_{i,j-1/2,k})  \simeq \frac{1}{h_y}\left(u(\bm{s}_{i,j,k}) - u(\bm{s}_{i,j-1,k}) \right)$\\
                             $\frac{\partial}{\partial z} u(\bm{s}_{i,j-1/2,k})  \simeq \frac{1}{4h_z}\left(u(\bm{s}_{i,j,k+1}) + u(\bm{s}_{i,j-1,k+1}) - u(\bm{s}_{i,j,k-1}) - u(\bm{s}_{i,j-1,k-1}) \right)$
                             \end{tabular}\\\hline          
        $\sigma_{i,j,k}^U$ & \begin{tabular}{l}
                             $\frac{\partial}{\partial x} u(\bm{s}_{i,j,k+1/2}) \simeq \frac{1}{4h_z}\left(u(\bm{s}_{i+1,j,k+1}) + u(\bm{s}_{i+1,j,k}) - u(\bm{s}_{i-1,j,k+1}) - u(\bm{s}_{i-1,j,k}) \right)$\\
                             $\frac{\partial}{\partial y} u(\bm{s}_{i,j,k+1/2})  \simeq \frac{1}{4h_y}\left(u(\bm{s}_{i,j+1,k+1}) + u(\bm{s}_{i,j+1,k}) - u(\bm{s}_{i,j-1,k+1}) - u(\bm{s}_{i,j-1,k}) \right)$\\
                             $\frac{\partial}{\partial z} u(\bm{s}_{i,j,k+1/2})  \simeq \frac{1}{h_x}\left(u(\bm{s}_{i,j,k+1}) - u(\bm{s}_{i,j,k}) \right) $
                             \end{tabular}\\\hline                  
        $\sigma_{i,j,k}^D$ & \begin{tabular}{l}
                             $\frac{\partial}{\partial x} u(\bm{s}_{i,j,k-1/2}) \simeq \frac{1}{4h_z}\left(u(\bm{s}_{i+1,j,k}) + u(\bm{s}_{i+1,j,k-1}) - u(\bm{s}_{i-1,j,k}) - u(\bm{s}_{i-1,j,k-1}) \right)$\\
                             $\frac{\partial}{\partial y} u(\bm{s}_{i,j,k-1/2})  \simeq \frac{1}{4h_y}\left(u(\bm{s}_{i,j+1,k}) + u(\bm{s}_{i,j+1,k-1}) - u(\bm{s}_{i,j-1,k}) - u(\bm{s}_{i,j-1,k-1}) \right)$\\
                             $\frac{\partial}{\partial z} u(\bm{s}_{i,j,k-1/2})  \simeq \frac{1}{h_x}\left(u(\bm{s}_{i,j,k}) - u(\bm{s}_{i,j,k-1}) \right)$
                             \end{tabular}\\\hline
    \end{tabular}
    \caption{Numerical scheme of the partial derivative with respect to $x$, $y$ and $z$ of $u_{ijk}$ on the different faces of cell $E_{ijk}$.}
    \label{tab:scheme}
\end{table}
\setlength\tabcolsep{6.0pt} 
Furthermore, let $\matr{H}$ be approximated by its value at the center of the face, and then, we have the approximation
\begin{equation}
    \begin{aligned}
        W^{\mathrm{dir}}_{ijk} =  & \int_{\sigma^{\mathrm{dir}}_{ijk}} \nabla u(\bm{s})^T \matr{H}(\bm{s})\bm{n}(\bm{s}) \mathrm{d}\sigma \\
        \approx & \nabla u(\bm{c}_{ijk}^\mathrm{dir})^T \matr{H}(\bm{c}_{ijk}^\mathrm{dir})\bm{n}(\bm{c}_{ijk}^\mathrm{dir}) \int_{\sigma^{\mathrm{dir}}_{ijk}} \mathrm{d}\sigma \\
        = & \nabla u(\bm{c}_{ijk}^\mathrm{dir})^T \matr{H}(\bm{c}_{ijk}^\mathrm{dir})\bm{n}(\bm{c}_{ijk}^\mathrm{dir})\mathrm{A}(\sigma^{\mathrm{dir}}_{ijk}),
    \end{aligned}
    \label{eq:faceint}
\end{equation}
where $\bm{c}_{ijk}^\mathrm{dir}$ is the center of face $\mathrm{dir}$ in the cell $E_{ijk}$, and $A(\sigma^{\mathrm{dir}}_{ijk})$ is the area of the face. 
Combining Equation~\eqref{eq:faceint} with the scheme of $\nabla u(\bm{c}_{ijk}^\mathrm{dir})$ from Table~\ref{tab:scheme}, and denoting the components of $\matr{H}$ as
\begin{equation*}
    \matr{H}(\bm{s}) = \begin{bmatrix}
    H^{11}(\bm{s}) & H^{12}(\bm{s}) & H^{13}(\bm{s}) \\
    H^{21}(\bm{s}) & H^{22}(\bm{s}) & H^{23}(\bm{s}) \\
    H^{31}(\bm{s}) & H^{32}(\bm{s}) & H^{33}(\bm{s}) 
    \end{bmatrix}
\end{equation*}
the approximations for each face become
\begin{equation*}
    \begin{aligned}
     &\hat{W}_{i,j,k}^R =\\ 
     & h_yh_z\left[\matr{H}^{11}(\bm{s}_{i+1/2,j,k})\frac{u(\bm{s}_{i+1,j,k}) - u(\bm{s}_{i,j,k})}{h_x}\right] + \\
     & h_yh_z\left[\matr{H}^{21}(\bm{s}_{i+1/2,j,k})\frac{u(\bm{s}_{i+1,j+1,k}) + u(\bm{s}_{i,j+1,k}) - u(\bm{s}_{i+1,j-1,k}) - u(\bm{s}_{i,j-1,k})}{4h_y} \right] + \\
     & h_yh_z\left[\matr{H}^{31}(\bm{s}_{i+1/2,j,k})\frac{u(\bm{s}_{i+1,j,k+1}) + u(\bm{s}_{i,j,k+1}) - u(\bm{s}_{i+1,j,k-1}) - u(\bm{s}_{i,j,k-1})}{4h_z} \right],
    \end{aligned}
\end{equation*}
\begin{equation*}
    \begin{aligned}
     &\hat{W}_{i,j,k}^L =\\ 
     & h_yh_z\left[\matr{H}^{11}(\bm{s}_{i-1/2,j,k})\frac{u(\bm{s}_{i-1,j,k}) - u(\bm{s}_{i,j,k})}{h_x}\right] +\\
     & h_yh_z\left[\matr{H}^{21}(\bm{s}_{i-1/2,j,k})\frac{u(\bm{s}_{i,j-1,k}) + u(\bm{s}_{i-1,j-1,k}) - u(\bm{s}_{i,j+1,k}) - u(\bm{s}_{i-1,j+1,k})}{4h_y} \right] + \\
     & h_yh_z\left[\matr{H}^{31}(\bm{s}_{i-1/2,j,k})\frac{u(\bm{s}_{i,j,k-1}) + u(\bm{s}_{i-1,j,k-1}) - u(\bm{s}_{i,j,k+1}) - u(\bm{s}_{i-1,j,k+1})}{4h_z} \right],
    \end{aligned}
\end{equation*}
\begin{equation*}
    \begin{aligned}
     &\hat{W}_{i,j,k}^B =\\ 
     & h_xh_z\left[\matr{H}^{12}(\bm{s}_{i,j+1/2,k})\frac{u(\bm{s}_{i+1,j+1,k}) + u(\bm{s}_{i+1,j,k}) - u(\bm{s}_{i-1,j+1,k}) - u(\bm{s}_{i-1,j,k})}{4h_x} \right] + \\
     & h_xh_z\left[\matr{H}^{22}(\bm{s}_{i,j+1/2,k})\frac{u(\bm{s}_{i,j+1,k}) - u(\bm{s}_{i,j,k})}{h_y}\right] + \\
     & h_xh_z\left[\matr{H}^{32}(\bm{s}_{i,j+1/2,k})\frac{u(\bm{s}_{i,j+1,k+1}) + u(\bm{s}_{i,j,k+1}) - u(\bm{s}_{i,j+1,k-1}) - u(\bm{s}_{i,j,k-1})}{4h_z} \right],
    \end{aligned}
\end{equation*}
\begin{equation*}
    \begin{aligned}
     &\hat{W}_{i,j,k}^F =\\ 
     & h_xh_z\left[\matr{H}^{12}(\bm{s}_{i,j-1/2,k})\frac{u(\bm{s}_{i-1,j,k}) + u(\bm{s}_{i-1,j-1,k}) - u(\bm{s}_{i+1,j,k}) - u(\bm{s}_{i+1,j-1,k})}{4h_x} \right]+\\
     & h_xh_z\left[\matr{H}^{22}(\bm{s}_{i,j-1/2,k})\frac{u(\bm{s}_{i,j-1,k}) - u(\bm{s}_{i,j,k})}{h_y}\right]+ \\
     & h_xh_z\left[\matr{H}^{32}(\bm{s}_{i,j-1/2,k})\frac{u(\bm{s}_{i,j,k-1}) + u(\bm{s}_{i,j-1,k-1}) - u(\bm{s}_{i,j,k+1}) - u(\bm{s}_{i,j-1,k+1})}{4h_z} \right],
    \end{aligned}
\end{equation*}
\begin{equation*}
    \begin{aligned}
     &\hat{W}_{i,j,k}^U =\\ 
     & h_xh_y\left[\matr{H}^{13}(\bm{s}_{i,j,k+1/2})\frac{u(\bm{s}_{i+1,j,k+1}) + u(\bm{s}_{i+1,j,k}) - u(\bm{s}_{i-1,j,k+1}) - u(\bm{s}_{i-1,j,k})}{4h_x} \right] +\\
     & h_xh_y\left[\matr{H}^{23}(\bm{s}_{i,j,k+1/2})\frac{u(\bm{s}_{i,j+1,k+1}) + u(\bm{s}_{i,j+1,k}) - u(\bm{s}_{i,j-1,k+1}) - u(\bm{s}_{i,j-1,k})}{4h_y} \right] +\\
     & h_xh_y\left[\matr{H}^{33}(\bm{s}_{i,j,k+1/2})\frac{u(\bm{s}_{i,j,k+1}) - u(\bm{s}_{i,j,k})}{h_z}\right],
    \end{aligned}
\end{equation*}
\begin{equation*}
    \begin{aligned}
     &\hat{W}_{i,j,k}^D =\\ 
     & h_xh_y\left[\matr{H}^{13}(\bm{s}_{i,j,k-1/2})\frac{u(\bm{s}_{i-1,j,k}) + u(\bm{s}_{i-1,j,k-1}) - u(\bm{s}_{i+1,j,k}) - u(\bm{s}_{i+1,j,k-1})}{4h_x} \right] + \\
     & h_xh_y\left[\matr{H}^{23}(\bm{s}_{i,j,k-1/2})\frac{u(\bm{s}_{i,j-1,k}) + u(\bm{s}_{i,j-1,k-1}) - u(\bm{s}_{i,j+1,k}) - u(\bm{s}_{i,j+1,k-1})}{4h_y} \right] + \\
     & h_xh_y\left[\matr{H}^{33}(\bm{s}_{i,j,k-1/2})\frac{u(\bm{s}_{i,j,k-1}) - u(\bm{s}_{i,j,k})}{h_z}\right],
    \end{aligned}
\end{equation*}
\begin{equation*}
    \begin{aligned}
     &\hat{W}_{i,j,k}^T =\\ 
     & h_xh_y\left[\matr{H}^{13}(\bm{s}_{i,j,k+1/2})\frac{u(\bm{s}_{i+1,j,k+1}) + u(\bm{s}_{i+1,j,k}) - u(\bm{s}_{i-1,j,k+1}) - u(\bm{s}_{i-1,j,k})}{4h_x} \right] +\\
     & h_xh_y\left[\matr{H}^{23}(\bm{s}_{i,j,k+1/2})\frac{u(\bm{s}_{i,j+1,k+1}) + u(\bm{s}_{i,j+1,k}) - u(\bm{s}_{i,j-1,k+1}) - u(\bm{s}_{i,j-1,k})}{4h_y} \right] +\\
     & h_xh_y\left[\matr{H}^{33}(\bm{s}_{i,j,k+1/2})\frac{u(\bm{s}_{i,j,k+1}) - u(\bm{s}_{i,j,k})}{h_z}\right],
    \end{aligned}
\end{equation*}
\begin{equation*}
    \begin{aligned}
     &\hat{W}_{i,j,k}^B =\\ 
     & h_xh_y\left[\matr{H}^{13}(\bm{s}_{i,j,k-1/2})\frac{u(\bm{s}_{i-1,j,k}) + u(\bm{s}_{i-1,j,k-1}) - u(\bm{s}_{i+1,j,k}) - u(\bm{s}_{i+1,j,k-1})}{4h_x} \right] + \\
     & h_xh_y\left[\matr{H}^{23}(\bm{s}_{i,j,k-1/2})\frac{u(\bm{s}_{i,j-1,k}) + u(\bm{s}_{i,j-1,k-1}) - u(\bm{s}_{i,j+1,k}) - u(\bm{s}_{i,j+1,k-1})}{4h_y} \right] + \\
     & h_xh_y\left[\matr{H}^{33}(\bm{s}_{i,j,k-1/2})\frac{u(\bm{s}_{i,j,k-1}) - u(\bm{s}_{i,j,k})}{h_z}\right].
    \end{aligned}
\end{equation*}

Next, a vectorization of the discretization is made; first moving along the $z$-direction, then along $x$-direction, and lastly along the $y$-direction. Let us denote this with the common index $l = j\cdot M\cdot P + i\cdot P + k$ so  $\bm{s}_{ijk}= \bm{s}_{j\cdot M\cdot P + i\cdot P + k} = \bm{s}_l$ which gives $u(\bm{s}_{ijk}) = u_l$ and $\kappa^2(\bm{s}_{ijk}) = \kappa_l^2$, and let the last index be $L = (N-1)MP + (M-1)P + P-1$. Further, the vectorization results in the linear system of equations
\begin{equation}
    (\matr{D}_V \matr{D}_{\kappa^2} - \matr{A}_H)\bm{u} = \matr{D}_V^{1/2}\bm{z},
    \label{eq:linrel}
\end{equation}
where $\matr{D}_V = V\cdot \matr{I}_{MNP}$, $\matr{D}_{\kappa^2} = \left[\kappa_0^2,\dots,\kappa_l^2,\dots,\kappa_{L}^2\right]\matr{I}_{MNP}$, and $\bm{z} \sim \norm(\bm{0},\matr{I}_{MNP})$. For simplicity the indices of the neighbors are denoted $k_p = k + 1$, $k_n = k - 1$, $j_p = j + 1$, $j_n = j - 1$, $i_p = i + 1$, and $i_n = i - 1$. The development of $\matr{A}_\matr{H}$ is done by the sum $\hat{W}^L_{ijk} + \hat{W}^R_{ijk} + \hat{W}^B_{ijk} + \hat{W}^F_{ijk} + \hat{W}^U_{ijk} + \hat{W}^D_{ijk}$ and accounting for the index in $u_{ijk}$ to form the linear relationship. 
In the following, non-zero elements of the $(jMN + iP + k)$-th row of $\matr{A}_\matr{H}$ are formalized, and the index in $(\matr{A}_\matr{H})_{\_}$ denotes the column being assigned. The resulting coefficient with the point itself is
\begin{equation*}
\begin{aligned}
    (\matr{A}_\matr{H})_{j\cdot M\cdot P + i\cdot P + k}  = & - \frac{h_yh_z}{h_x}\left[ \matr{H}^{11}(\bm{s}_{i+1/2,j,k})+\matr{H}^{11}(\bm{s}_{i-1/2,j,k})\right] \\
    & - \frac{h_xh_z}{h_y}\left[ \matr{H}^{22}(\bm{s}_{i,j+1/2,k})+\matr{H}^{22}(\bm{s}_{i,j-1/2,k})\right] \\
    & - \frac{h_xh_y}{h_z}\left[ \matr{H}^{33}(\bm{s}_{i,j,k+1/2})+\matr{H}^{22}(\bm{s}_{i,j,k-1/2})\right],
\end{aligned}
\end{equation*}
with the six closest neighbors are
\begin{equation*}
\begin{aligned}
    (\matr{A}_H)_{j\cdot M\cdot P + i\cdot P + k_p}  = &  \frac{h_xh_y}{h_z}\matr{H}^{33}(\bm{s}_{i,j,k+1/2}) \\
    & + \frac{h_y}{4}\left[\matr{H}^{31}(\bm{s}_{i+1/2,j,k})  - \matr{H}^{31}(\bm{s}_{i-1/2,j,k})\right] \\
    & + \frac{h_x}{4}\left[\matr{H}^{32}(\bm{s}_{i,j+1/2,k})  - \matr{H}^{32}(\bm{s}_{i,j-1/2,k}) \right]
\end{aligned}
\end{equation*}
\begin{equation*}
\begin{aligned}
    (\matr{A}_H)_{j\cdot M\cdot P + i\cdot P + k_n}  = &  \frac{h_xh_y}{h_z}\matr{H}^{33}(\bm{s}_{i,j,k-1/2}) \\
    & - \frac{h_y}{4}\left[\matr{H}^{31}(\bm{s}_{i+1/2,j,k})  - \matr{H}^{31}(\bm{s}_{i-1/2,j,k})\right] \\
    & - \frac{h_x}{4}\left[\matr{H}^{32}(\bm{s}_{i,j+1/2,k})  - \matr{H}^{32}(\bm{s}_{i,j-1/2,k})\right]
\end{aligned}
\end{equation*}
\begin{equation*}
\begin{aligned}
    (\matr{A}_H)_{j\cdot M\cdot P + i_p\cdot P + k}  = & \frac{h_zh_y}{h_x}\matr{H}^{11}(\bm{s}_{i+1/2,j,k}) \\
    & + \frac{h_y}{4}\left[\matr{H}^{12}(\bm{s}_{i,j,k+1/2})  - \matr{H}^{12}(\bm{s}_{i,j,k-1/2})\right] \\
    & + \frac{h_z}{4}\left[\matr{H}^{13}(\bm{s}_{i,j+1/2,k})  - \matr{H}^{13}(\bm{s}_{i,j-1/2,k}) \right]
\end{aligned}
\end{equation*}
\begin{equation*}
\begin{aligned}
    (\matr{A}_H)_{j\cdot M\cdot P + i_n\cdot P + k} = & \frac{h_zh_y}{h_x}\matr{H}^{11}(\bm{s}_{i-1/2,j,k}) \\
    & - \frac{h_y}{4}\left[\matr{H}^{12}(\bm{s}_{i,j,k+1/2})  - \matr{H}^{12}(\bm{s}_{i,j,k-1/2})\right] \\
    & - \frac{h_z}{4}\left[\matr{H}^{13}(\bm{s}_{i,j+1/2,k})  - \matr{H}^{13}(\bm{s}_{i,j-1/2,k}) \right]
\end{aligned}
\end{equation*}
\begin{equation*}
\begin{aligned}
    (\matr{A}_H)_{j_p\cdot M\cdot P + i\cdot P + k} = & \frac{h_xh_z}{h_y}\matr{H}^{22}(\bm{s}_{i,j+1/2,k}) \\
    & + \frac{h_x}{4}\left[\matr{H}^{23}(\bm{s}_{i,j,k+1/2})  - \matr{H}^{23}(\bm{s}_{i,j,k-1/2})\right] \\
    & + \frac{h_z}{4}\left[\matr{H}^{21}(\bm{s}_{i+1/2,j,k})  - \matr{H}^{21}(\bm{s}_{i-1/2,j,k}) \right]
\end{aligned}
\end{equation*}
\begin{equation*}
\begin{aligned}
    (\matr{A}_H)_{j_n\cdot M\cdot P + i\cdot P + k}  = & \frac{h_xh_z}{h_y}\matr{H}^{22}(\bm{s}_{i,j-1/2,k}) \\
    & - \frac{h_x}{4}\left[\matr{H}^{23}(\bm{s}_{i,j,k+1/2})  - \matr{H}^{23}(\bm{s}_{i,j,k-1/2})\right] \\
    & - \frac{h_z}{4}\left[\matr{H}^{21}(\bm{s}_{i+1/2,j,k})  - \matr{H}^{21}(\bm{s}_{i-1/2,j,k}) \right],
\end{aligned}
\end{equation*}
and with the twelve closest diagonals are
\begin{equation*}
    (\matr{A}_H)_{j\cdot M\cdot P + i_p\cdot P + k_p} = \frac{h_y}{4}\left[\matr{H}^{31}(\bm{s}_{i+1/2,j,k}) + \matr{H}^{13}(\bm{s}_{i,j,k+1/2}) \right],
\end{equation*}
\begin{equation*}
    (\matr{A}_H)_{j\cdot M\cdot P + i_n\cdot P + k_n} = \frac{h_y}{4}\left[\matr{H}^{31}(\bm{s}_{i-1/2,j,k}) + \matr{H}^{13}(\bm{s}_{i,j,k-1/2}) \right],
\end{equation*}
\begin{equation*}
    (\matr{A}_H)_{j\cdot M\cdot P + i_n\cdot P + k_p} = - \frac{h_y}{4}\left[ \matr{H}^{31}(\bm{s}_{i-1/2,j,k}) + \matr{H}^{13}(\bm{s}_{i,j,k+1/2}) \right]
\end{equation*}
\begin{equation*}
    (\matr{A}_H)_{j\cdot M\cdot P + i_p\cdot P + k_n} = - \frac{h_y}{4}\left[\matr{H}^{31}(\bm{s}_{i+1/2,j,k}) + \matr{H}^{13}(\bm{s}_{i,j,k-1/2}) \right],
\end{equation*}
\begin{equation*}
    (\matr{A}_H)_{j_p\cdot M\cdot P + i\cdot P + k_p} = \frac{h_x}{4}\left[\matr{H}^{32}(\bm{s}_{i,j+1/2,k}) + \matr{H}^{23}(\bm{s}_{i,j,k+1/2}) \right],
\end{equation*}
\begin{equation*}
    (\matr{A}_H)_{j_n\cdot M\cdot P + i\cdot P + k_n} = \frac{h_x}{4}\left[\matr{H}^{32}(\bm{s}_{i,j-1/2,k}) + \matr{H}^{23}(\bm{s}_{i,j,k-1/2}) \right],
\end{equation*}
\begin{equation*}
    (\matr{A}_H)_{j_n\cdot M\cdot P + i\cdot P + k_p} = - \frac{h_x}{4}\left[\matr{H}^{32}(\bm{s}_{i,j-1/2,k}) + \matr{H}^{23}(\bm{s}_{i,j,k+1/2}) \right],
\end{equation*}
\begin{equation*}
    (\matr{A}_H)_{j_p\cdot M\cdot P + i\cdot P + k_n} = - \frac{h_x}{4}\left[\matr{H}^{32}(\bm{s}_{i,j+1/2,k}) + \matr{H}^{23}(\bm{s}_{i,j,k-1/2}) \right],
\end{equation*}
\begin{equation*}
    (\matr{A}_H)_{j_p\cdot M\cdot P + i_p\cdot P + k} = \frac{h_z}{4}\left[\matr{H}^{21}(\bm{s}_{i+1/2,j,k}) + \matr{H}^{12}(\bm{s}_{i,j+1/2,k}) \right],
\end{equation*}
\begin{equation*}
    (\matr{A}_H)_{j_n\cdot M\cdot P + i_n\cdot P + k} = \frac{h_z}{4}\left[\matr{H}^{21}(\bm{s}_{i-1/2,j,k}) + \matr{H}^{12}(\bm{s}_{i,j-1/2,k}) \right],
\end{equation*}
\begin{equation*}
    (\matr{A}_H)_{j_n\cdot M\cdot P + i_p\cdot P + k} = - \frac{h_z}{4}\left[\matr{H}^{21}(\bm{s}_{i+1/2,j,k}) + \matr{H}^{12}(\bm{s}_{i,j-1/2,k}) \right],
\end{equation*}
\begin{equation*}
    (\matr{A}_H)_{j_p\cdot M\cdot P + i_n\cdot P + k} = - \frac{h_z}{4}\left[\matr{H}^{21}(\bm{s}_{i-1/2,j,k}) + \matr{H}^{12}(\bm{s}_{i,j+1/2,k}) \right].
\end{equation*}
Note that the corner points are not included in this scheme. 
Denoting $\matr{A} = \matr{D}_V\matr{D}_{\kappa^2} - \matr{A}_\matr{H}$, Equation~\eqref{eq:linrel} can be written as
\begin{equation*}
    \bm{z} = \matr{D}_V^{-1/2} \matr{A}\bm{u},
\end{equation*}
and thus, the joint distribution of $\bm{u}$ is
\begin{equation*}
    \pi(\bm{u}) \propto \pi(\bm{z}) \propto \exp\left( -\frac{1}{2} \bm{z}^\mathrm{T}\bm{z}\right)
\end{equation*}
\begin{equation*}
    \pi(\bm{u}) \propto \exp\left( -\frac{1}{2} \bm{u}^\mathrm{T}\matr{A}^\mathrm{T}\matr{D}_V^{-1}\matr{A}\bm{u}\right)
\end{equation*}
\begin{equation*}
    \pi(\bm{u}) \propto \exp\left( -\frac{1}{2} \bm{u}^\mathrm{T}\matr{Q}\bm{u}\right).
\end{equation*}
Here, $\matr{Q} = \matr{A}^\mathrm{T}\matr{D}_V^{-1}\matr{A}$ which is a sparse matrix of $93$ non-zero elements per row. This corresponds to the point, the 18 closest neighbors, and their 18 closest neighbors. Then removing duplicates results in 93 points. 

\section{Additional figures}
\label{sec:addfig}

\begin{figure}[!ht]
    \centering
    \begin{subfigure}[b]{0.3\textwidth}
         \centering
         \includegraphics[width=\textwidth]{figures/meanApp.png}
         \caption{SINMOD prior}
         \label{fig:prior}
    \end{subfigure}
    \begin{subfigure}[b]{0.3\textwidth}
         \centering
         \includegraphics[width=\textwidth]{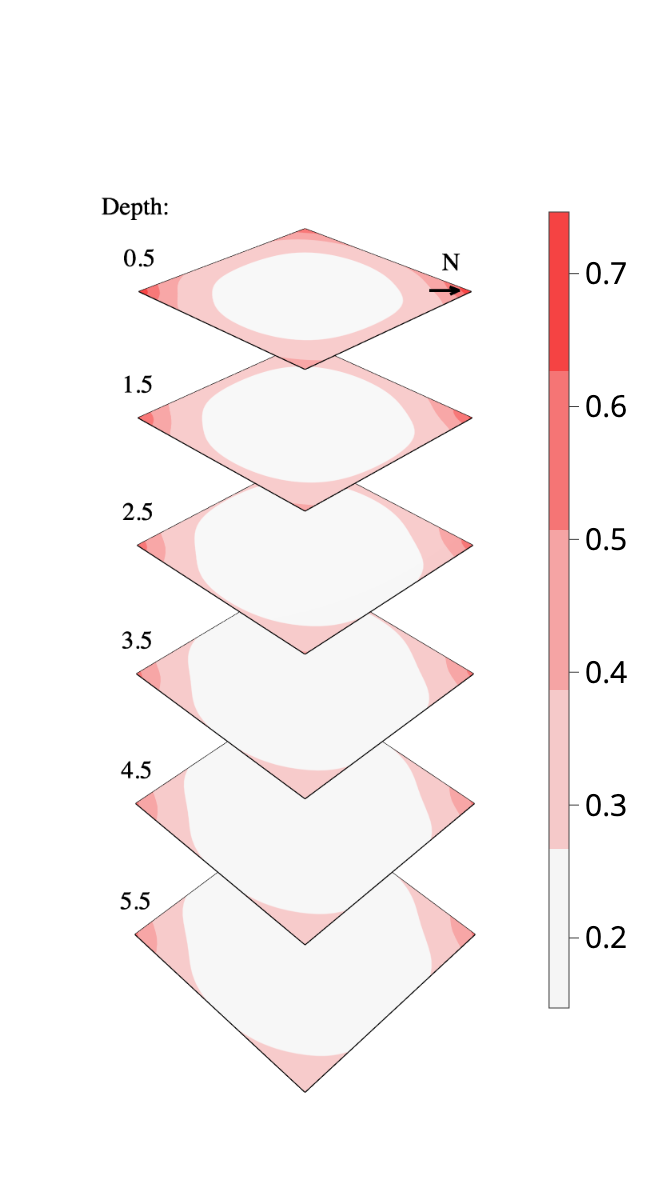}
         \caption{Marginal Variance}
         \label{fig:SAmvar}
    \end{subfigure}
    \begin{subfigure}[b]{0.3\textwidth}
         \centering
         \includegraphics[width=\textwidth]{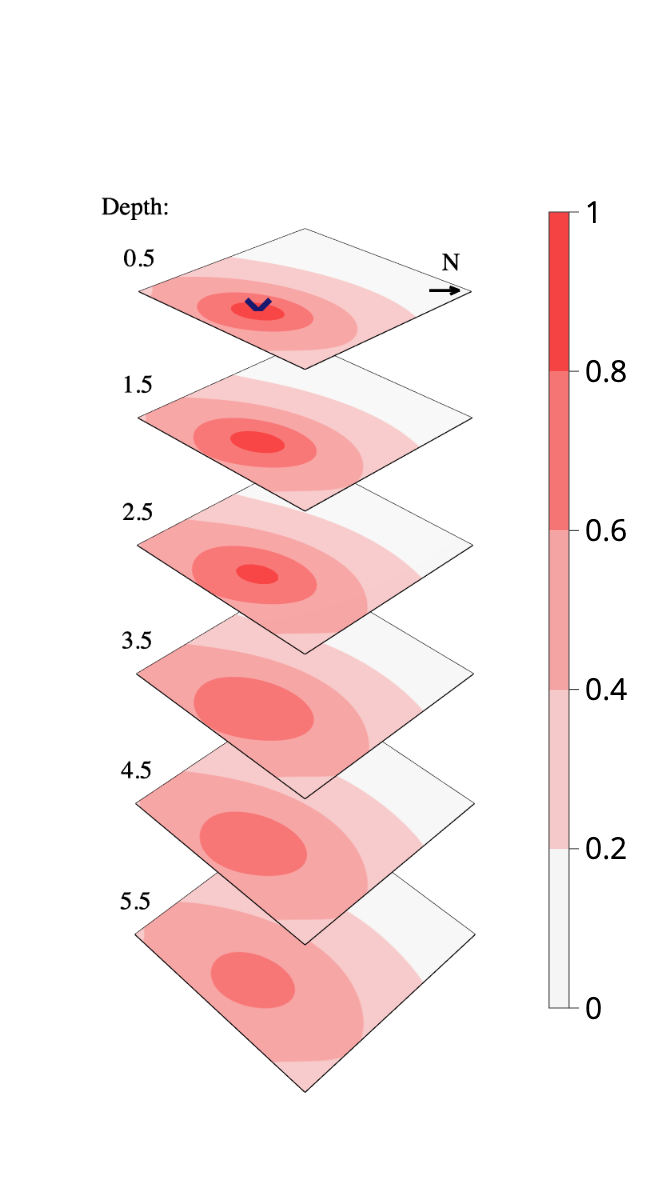}
         \caption{Correlation}
         \label{fig:SAcorr}
    \end{subfigure}
    \caption{\label{fig:SAapp} Prior field \textbf{(a)} found from SINMOD simulations, the variance of the spatial effect \textbf{(b)} and spatial correlation of point [22,10,0] \textbf{(c)} in the stationary anisotropic model. The N-arrow shows the cardinal north.}
\end{figure}

In the application, Section~5, we estimate the parameters of a non-stationary anisotropic and stationary anisotropic model on a simulated dataset from the numerical ocean model SINMOD. The resulting properties of the non-stationary model are presented in Figure~7 in Section~5.2 since this is the main focus of the applications. The properties of the stationary anisotropic model fit on the same dataset are presented in Figure~\ref{fig:SAapp}. The marginal variance in Figure~\ref{fig:SAmvar}, which should be constant for this stationary model, shows some variability caused by the boundary conditions. Notice that this boundary effect is also bigger in the direction of the strongest dependency directions seen in the south and north corners. Notice also the large discrepancies between the correlations in these two models, Figure~\ref{fig:SAcorr} and Figure~7c, as the stationary anisotropic model kind of captures an average correlation within the field.

\newpage
\bibliographystyle{apalike}
\bibliography{main}
\end{document}